\documentclass[twocolumn,showpacs,preprintnumbers,amsmath,amssymb]{revtex4}
\usepackage{pifont}

\usepackage{graphicx}
\usepackage{dcolumn}
\usepackage{bm}
\usepackage[colorlinks,linkcolor=blue,anchorcolor=blue,citecolor=blue]{hyperref}
\usepackage{lineno} 
\usepackage{amssymb}
\usepackage{amsmath}
\usepackage{latexsym}
\usepackage{mathrsfs}

\begin{document}
\preprint{}

\title{L\'{e}vy walk dynamics in non-static media}

\author{Tian Zhou$^{1}$}
\author{Pengbo Xu$^{2}$}
\author{Weihua Deng$^{1}$}
\affiliation{$^1$School of Mathematics and Statistics, Gansu Key Laboratory of Applied Mathematics and Complex Systems, Lanzhou University, Lanzhou 730000, P.R. China}
\affiliation{$^2$School of Mathematical Sciences, Peking University, Beijing 100871, P.R. China}


\begin{abstract}

Almost all the media the particles move in are non-static. Depending on the expected resolution of the studied dynamics and the amplitude of the displacement of the media, sometimes the non-static behaviours of the media can not be ignored.  In this paper, we build the model describing L\'evy walks in non-static media, where the physical and comoving coordinates are connected by scale factor. We derive the equation governing the probability density function of the position of the particles in comoving coordinate. Using the Hermite orthogonal polynomial expansions, some statistical properties are obtained, such as mean squared displacements (MSDs) in both coordinates and kurtosis. For some representative non-static media and L\'{e}vy walks, the asymptotic behaviors of MSDs in both coordinates are analyzed in detail. The stationary distributions and mean first passage time for some cases are also discussed through numerical simulations.


\end{abstract}

\maketitle

\section{Introduction}

The diffusion processes can be classified according to the relation between the mean squared displacement (MSD) defined as $\langle x^2(t)\rangle=\int_{-\infty}^{\infty} x^2P(x,t)dx$ and the time $t$. Usually, the MSD appears as $\langle x^2(t)\rangle\sim t^\alpha$; the diffusion process is called normal diffusion if $\alpha=1$, otherwise it is termed as anomalous diffusion \cite{BG1990,SLLS2004,MJCB2014}. A representative example of normal diffusion is Brownian motion \cite{NelsonE}, which is discovered by irregular motion of the pollen and other suspended particles in water. Anomalous diffusion is ubiquitous in a wide range of fields, such as diffusion processes in physics \cite{Norregaard2017}, in finance \cite{s3}, in ecology \cite{s4}, and in biology \cite{Franosch2013}, etc. Specifically, an anomalous diffusion is called subdiffusion \cite{SM1975,SW2009,JLOM2013,WSTK2011,JMJM2012,WST2010,TBKK2013} if $0<\alpha<1$ and superdiffusion \cite{CGE2000,BS2002,RJLM2015} if $\alpha>1$. Random walk process, including continuous time random walk (CTRW) and L\'{e}vy walk, is one of the most popular models to describe anomalous diffusion. There are two series of independent identically distributed (i.i.d.) random variables in CTRW model, which respectively are waiting time~$\tau$~following the distribution $\phi(\tau)$ and jump length~$l$~following the probability density function (PDF) $\lambda(l)$ \cite{s16,s11,M1969}. The corresponding subdiffusion or superdiffusion process can be modeled with infinite $\langle \tau\rangle$ and finite $\langle l^2\rangle$ or finite $\langle \tau\rangle$ and infinite $\langle l^2\rangle$, respectively. L\'evy flight \cite{F1994a,F1994b,s16} is a classical model which displays superdiffusion when the jump length density behaves as $\lambda(l)\sim1/|l|^{1+\mu}$, $0<\mu<2$. L\'{e}vy walk is another important model of random walks which is space and time coupled through finite propagation speed \cite{Zaburdaev,BFK2000}. The traditional L\'{e}vy walk is the one with constant value of speed $v_0$ and the corresponding walking length of which for each finished step is $v_0 \tau$, where $\tau$ is the walking time for each step of movement following the density $\phi(\tau)$.

The ordinary diffusion process mentioned above is implicitly with the assumption that the medium is static, which implies the distance between two stationary particles does not change with time. In reality, almost all the media the particles move in are non-static. The expected resolution of the studied dynamics and the amplitude of the displacement of the media decide whether or not the non-static behaviors can be ignored.  The expansion and contraction are two typical features of the media, which 
can be observed in many different areas, such as biology \cite{Simpson2015,Simpson2015b,Crampin1999,Crampin2002}, cosmology \cite{Berezinsky2006,Berezinsky2007,Aloisio2009}, fluids \cite{Haba2014}. In biology, the formation of pigmentation patterns depends on the growing concomitant tissues and organs. High energy cosmic rays is also an example of expansion from the perspective of cosmology \cite{Kotera2008}. Moreover, in \cite{Crampin1999,Berezinsky2006,Yates2014,Averbukh2014,Simpson2015,Simpson2015b,Yuste2016,F}, the authors show that the expansion of medium brings about enormous difference on the behavior of the diffusion particles, which motivates people to pay attention to the movement of the particles in non-static medium.   

A growing number of scholars concentrate on analyzing the transport characters of diffusion process in non-static medium. The relevant Fokker-Planck equation which describes diffusion in non-static medium has been derived via generalized Chapman-Kolmogorov equation in \cite{Yuste2016}. The crossover effects has been discovered in a non-static medium with a power-law scale factor. Moreover, in \cite{F}, the authors introduce a comoving coordinate which can be thought as a reference frame where the medium looks to be static. Further in \cite{F,Sokolov2012}, the Fokker-Planck equation in comoving coordiante as well as the long-time asymptotic behavior of comoving MSDs have been derived for CTRW model in a non-static medium. Additionally, based on the relation between physical and comoving coordinates, the relevant physical statistics of CTRW in a non-static medium are obtained. Moreover, by introducing the conformal time, the dynamical behavior of L\'{e}vy flight in a non-static medium has also been discussed in \cite{F}, which concludes that in a non-static medium the diffusion coefficient of fractional diffusion equation for subdiffusive CTRW is time-dependent.

In this paper we discuss the dynamical behavior of L\'{e}vy walk in non-static medium. We first build the model and derive the governing equations. Then using Hermite polynomials expansion established in \cite{Pengbo} solves the equation to calculate the statistical observables. 
The method of Hermite orthogonal polynomials expansion is good at calculating MSD even for some cases that are unsolvable by ordinary method of integral transform, such as L\'evy walks moving under external potential \cite{Pengbo1,Tian2021}. In addition, for the Langevin equation describing the diffusion processes under the action of external potential \cite{SIM2003,coffey,Metzle1999,JMF1999,CKGM2003}, Hermite polynomials expansion is also a doable method 
\cite{Barkai1998,Metzle1999,Henry2010,Langlands2010}.

 This paper is organized as follows. In Sec. \ref{sec 2}, we introduce the L\'{e}vy walk model in a non-static medium and construct the master equation in the frame of comoving coordinate. In Sec. \ref{sec 3}, the corresponding MSDs are calculated in comoving coordinate by the approach of Hermite polynomials expansion, and the asymptotic behaviors of MSDs in physical space are obtained as well. In Sec. \ref{sec 4}, the kurtosis in physical coordinate for the case of localization with exponential scale factor is calculated and the evolutions are analyzed in detail; the stationary distributions for some cases are also discussed through numerical simulations. In Sec.  \ref{sec 5}, we obtain the mean first passage time by numerical simulations. The paper is concluded with some discussions in Sec. \ref{sec 6}.

\section{L\'evy walk in a non-static media}\label{sec 2}


The most representative L\'evy walk process is isotropic and with constant velocity $v_0$ \cite{Zaburdaev,First}, 
and then the length that L\'evy walk particle moves for each step of renewal is $v_0 \tau$, where $\tau$ indicating the walking duration of each step follows the density $\phi(\tau)$. If L\'evy walk particle stays at position $x$ after finishing some steps of moving, then the next step will arrive at $x \pm v_0 \tau$. It can be seen that the spatiotemporal coupling is 
the most striking feature of L\'evy walk.  Another significant difference between L\'evy walk and CTRW is that the former one has no rest after each step of walking, while the latter one must rest after each instant jumping, leading to much more difference when considering non-static media.

%

In this paper, we focus on one-dimension system with uniform expansion or contraction, and the physical coordinate (denoted as $y$) and comoving coordinate (denoted as $x$) are correlated with each other in the following way
\begin{equation}\label{1.2}
  y(t)=a(t)x,
\end{equation}
where $a(t_0)=1$ for the initial time $t_0$. The function $a(t)$ is called as scale factor in the context of
cosmology \cite{cosmological,Introducing}. Since the initial condition of $a(t_0)=1$,  the physical coordinate and comoving coordinate are same at time $t_0$. According to the property of scale factor $a(t)$, we can classify the medium; specifically, we say a medium is expanding (or contracting) if $\dot{a}(t)>0$ (or $\dot{a}(t)<0$). One can see that the distance between two stationary particles always changes with time, since the medium the particles are embedded always expands or contracts. 

\begin{figure}[htbp]
\centering
\includegraphics[scale=0.45]{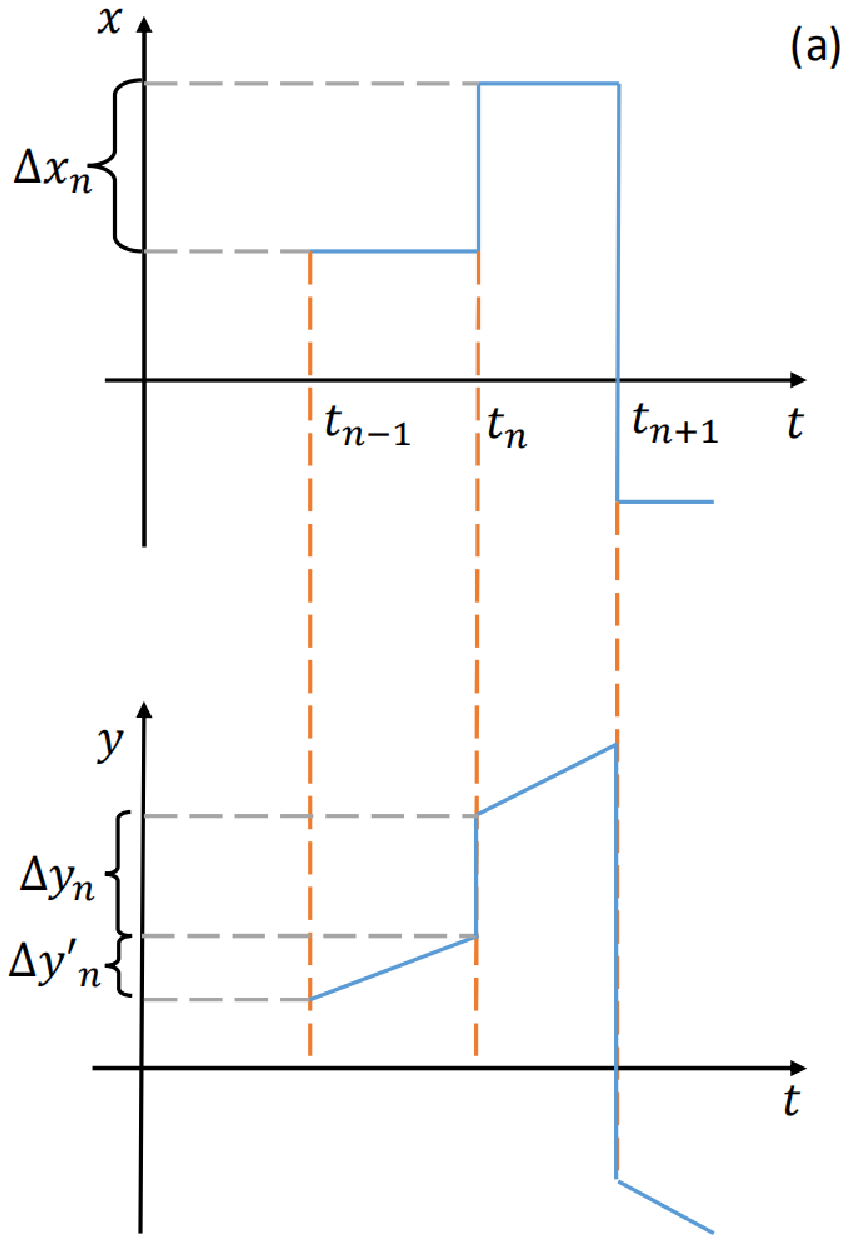}
\includegraphics[scale=0.45]{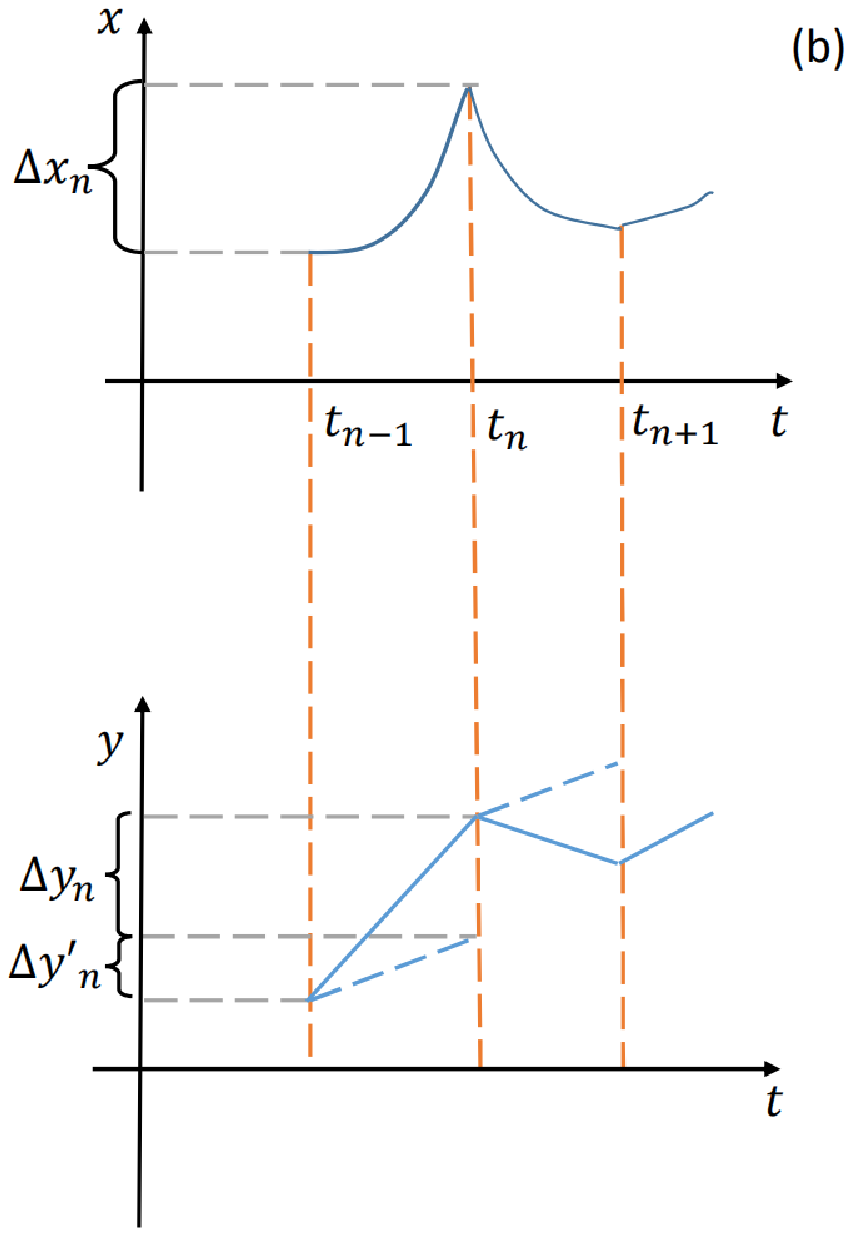}
\caption{CTRW and L\'evy walk in expanding media. The comoving and physical coordinates are $x$ and $y$, respectively. Figure (a) represents CTRW process in expanding medium, where $\Delta y_n$ is a random variable drawn from the PDF $\lambda(\Delta y_n)$, $\Delta y'_n$ represents the displacement caused by expanding medium from time $t_{n-1}$ to $t_n$, and $\Delta x_n=\Delta y_n/a(t_n)$ denotes the jumping length in comoving space. Figure (b) illustrates L\'evy walk process in expanding medium with $\Delta y_n = v_0 (t_n-t_{n-1})$, and the other notations share similar meanings with those in (a).}
\label{model}
\end{figure}

The CTRW model in a non-static medium has been studied in \cite{F}. Let us take the $n$th step of CTRW process for instance. For ordinary CTRW model, the process will wait at $y_{n-1}$ for a period of random time then instantly jumps $\Delta y_n$ relative to $y_{n-1}$ drawn from the PDF $\lambda(\Delta y_n)$. However, the process will move along with the non-static medium, even though it is waiting for next jump as shown in Fig. \ref{model}(a). Assuming the CTRW particle arrives at $y_{n-1}$ at time $t_{n-1}$, then due to non-static medium, it comes to $y'_{n-1}$ at time $t_n$, and further has a jump with the length of $\Delta y_n$ relative to $y'_{n-1}$. Since the scale factors also change with time, we must borrow comoving coordinates to determine $y'_{n-1}$, which is  $x_{n-1}^+=x_n^-$ (`+' and `-' respectively mean right and left limit); further from the relation between physical coordinate and comoving coordinate \eqref{1.2}, it holds
\begin{equation*}
	y'_{n-1}= \frac{a(t_n)}{a(t_{n-1})} y_{n-1}.
\end{equation*}
We denote $\Delta y'_n=y'_{n-1}-y_{n-1}$.

In this paper, we mainly consider the dynamic behavior of L\'evy walk in a non-static medium. As shown in Fig. \ref{model}(b), since L\'evy walk process has no rest, the trajectory behaves much different. Consider the time $\xi \in (t_{n-1},t_n]$. Since L\'evy walk does not wait, its  movement in physical coordinate consists of two parts: one is movement caused by non-static medium,  $y'_\xi = (a(\xi)/a(t_{n-1})) y_{n-1}$; the other one is walking length relative to $y'_\xi$, that is, $\pm v_0 (\xi-t_{n-1})$. Whereas in CTRW model, if we also consider the time $\xi$ between the $(n-1)$th and $n$th jumping events, then the movement is only influenced by non-static medium. This may cause a major difference between CTRW and L\'evy walk models, in fact, assuming that $t=\xi$ is the final observation time, then at time $t$ the L\'evy walk particle in non-static medium locates at
\begin{equation}\label{model_yt}
	y_{t}= \frac{a(t)}{a(t_{n})} y_{n} \pm v_0 (t-t_n).
\end{equation}
Obviously, the static media whose scale factor $a(t)=1$ is the case considered in ordinary L\'evy walk problem  \cite{Zaburdaev,First}, which indicates the distance between two stationary particles does not change with time.

As discussed above, the walking length $\Delta y$ for L\'evy walk in physical coordinate is determined by the walking time $\tau$ drawn from the PDF $\phi(\tau)$ or the survival time; with a little abuse of notation we still denote it as $\tau$ following the survival probability
\begin{equation}\label{survival_prob_t}
	\psi(\tau)=\int_\tau^\infty \phi(\tau') d \tau',
\end{equation}
therefore $\Delta y$ can be given by the following conditional density
\begin{equation}\label{1.3}
 \lambda(\Delta y \vert \tau)=\frac{1}{2}\delta(\Delta y-v_0\tau)+\frac{1}{2}\delta(\Delta y+v_0\tau),
\end{equation}
where $\delta(\cdot)$ represents the Dirac $\delta$-function.
It is not easy to directly consider the process in physical coordinate since the relation of \eqref{model_yt} is hard to be conveniently utilized. Therefore in the following we will change the process into the one with comoving coordinate.

%

In order to transfer the L\'evy walk process from physical coordinate to comoving coordinate, first of all we need to identify the walking length $\Delta x$ for each step of moving. It can be accomplished by the relation between $\Delta y$ and $\Delta x$ at time $t$ \cite{F}, i.e, $\Delta y=a(t) \Delta x$; specially we choose $t$ to be the renewal time when the walker finishes a few steps of walking (such as $n$ steps) or the overall observation time, and take $\tau$ to be the walking time of current step (the $n$th step) or the survival time so that $(t-\tau)$ should be the time when current step starts. Denote the conditional density of $\Delta x$ as $\lambda(\Delta x \vert t; \tau)$, which means the particle in comoving coordinate walks $\Delta x$ for a period of $\tau$ at time $t$.
%
According to the relation between $\Delta y$ and $\Delta x$ and combining with \eqref{1.3}, one has
\begin{equation}\label{1.4}
\begin{split}
   \lambda(\Delta &x|t;\tau) = a(t) \lambda\big(a(t) \Delta x \vert \tau\big) \\
     & =\frac{1}{2}a(t)\left( \delta \left(a(t) \Delta x-v_0 \tau\right)+\delta \left(a(t) \Delta x+v_0 \tau\right)\right)\\
     &=\frac{1}{2} \left[ \delta\left( \Delta x -\frac{v_0\tau}{a(t)} \right)+ \delta\left( \Delta x +\frac{v_0\tau}{a(t)} \right)\right].
\end{split}
\end{equation}

We use $W(x,t)$ to denote the PDF of finding a particle at position $x$ in comoving coordinate at time $t$. Assume the related PDF in physical space to be $P(y,t)$. According to the relation between comoving and physical coordinates \eqref{1.2}, one gets
\begin{equation}\label{1.5}
  P(y,t)= \frac{1}{a(t)} W\left(\frac{y}{a(t)},t\right).
\end{equation}
The PDF $P(y,t)$ as well as some important statistics such as the first two moments of L\'evy walk in non-static medium can be calculated according to \eqref{1.5}. In the following we aim to derive the transport equations of L\'{e}vy walk in comoving coordinate. As one can see in the comoving coordinate, if the particle locates at $x$ at time $t$ which is the renewal time, and the walking time for this step $\tau$ following the PDF $\phi(\tau)$, then the previous renewal event happens at time $t-\tau$ at position $x - \Delta x$ with $\Delta x$ given by $\lambda(\Delta x \vert t; \tau)$. If we denote the corresponding PDF of finding particle just arriving at position $x$ at renewal time $t$ as $q(x,t)$, accordingly we have
\begin{equation}\label{1}
\begin{split}
   q(x,t) & \! =\!\int_{-\infty}^{\infty}d\Delta x\int_{0}^{t} q\Big(x-\Delta x,t-\tau\Big)\phi(\tau)\lambda(\Delta x|t;\tau)d\tau\\
     & +P_0(x)\delta(t),
\end{split}
\end{equation}
where the second term on r.h.s. represents the initial distribution.
Then the PDF in the comoving coordinate denoted as $W(x,t)$  can be given though \eqref{1}, which is
\begin{equation}\label{1.}
   W(x,t)\!=\!\int_{-\infty}^{\infty}d\Delta x\int_{0}^{t} q\Big(x-\Delta x,t-\tau\Big)\psi(\tau)\lambda(\Delta x|t;\tau)d\tau,
\end{equation}
where $\psi(\tau)$ is defined in \eqref{survival_prob_t}. By taking Laplace transform, defined as $\hat{g}(s)=\mathscr{L}_t \{g(t)\}(s)=\int_{0}^{\infty} e^{-s t}g(t)dt$, on survival probability $\psi(\tau)$ there exists
\begin{equation}\label{survival_prob}
  \hat{\psi}(s)=\frac{1-\hat{\phi}(s)}{s}.
\end{equation}
%
Substituting the expression of $\lambda(\Delta x|t;\tau)$ given by \eqref{1.4} into \eqref{1} and \eqref{1.}, respectively, we have
%
%
%
\begin{equation}\label{1..6}
\begin{split}
    q(x,&t)=\frac{1}{2}\int_{0}^{t}q\bigg(x-\frac{v_0\tau}{a(t)},t-\tau\bigg)\phi(\tau)d\tau  \\
     &+\frac{1}{2} \int_{0}^{t}q\bigg(x+\frac{v_0\tau}{a(t)},t-\tau\bigg)\phi(\tau)d\tau +P_0(x)\delta(t)
\end{split}
\end{equation}
and
\begin{equation}\label{1..7}
\begin{split}
    W(x,t)&=\frac{1}{2} \int_{0}^{t}q\bigg(x-\frac{v_0\tau}{a(t)},t-\tau\bigg)\psi(\tau)d\tau  \\
     &+\frac{1}{2} \int_{0}^{t}q\bigg(x+\frac{v_0\tau}{a(t)},t-\tau\!\bigg)\psi(\tau)d\tau.
\end{split}
\end{equation}
%
The normalization of $W(x,t)$ can be immediately verified by Fourier transform, defined as $\tilde{f}(k)=\mathscr{F}_{x}\{f(x)\}(k)=\int_{-\infty}^{\infty}e^{-i k x}f(x)dx$. In fact by taking Fourier-Laplace transform on \eqref{1..6} and \eqref{1..7}, and letting $k=0$, there exists
\begin{equation*}
\begin{split}
  \hat{\tilde{q}}(0,s) &=\frac{1}{1-\hat{\phi}(s)},\\
  \widehat{\widetilde{W}}(0,s) &= \hat{\psi}(s) \hat{\tilde{q}}(0,s)=1/s,
\end{split}
\end{equation*}
indicating normalization of PDF $W(x,t)$ after inverse Laplace transform.

In the following, to theoretically obtain the behaviors of moments in comoving and physical coordinates, we utilize Hermite orthogonal polynomials to approach PDF \cite{Pengbo1,Pengbo,Tian2021}.
%
%

\section{Hermite polynomial approximation to L\'{e}vy walk in non-static medium}\label{sec 3}

The Hermite polynomials form an orthogonal basis of the Hilbert space with the inner product $\langle f,g \rangle=\int_{-\infty}^{\infty} f(x)\bar{g}(x)e^{-x^2} dx$ \cite{hermit_intro}. In this section, we utilize the Hermite polynomials to approach the PDF of L\'{e}vy walk in comoving coordinate. According to the complete orthogonal system, we assume that in Hilbert space $q(x,t)$ and $W(x,t)$ can be, respectively, represented as
\begin{align}
	q(x,t)&=\sum_{n=0}^{\infty} H_n(x) T_n(t) e^{-x^2},\label{1.9}\\
	W(x,t)&=\sum_{n=0}^{\infty} H_n(x) R_n(t) e^{-x^2},\label{1.10}
\end{align}
where $H_n(x), n=0,1,\cdots,$ represent the Hermite polynomials, $\{T_n(t)\}$ and $\{R_n(t)\}$ are a series of functions with respect to $t$ to be determined.

We take the initial distribution of the particle as a Dirac $\delta$-function, i.e., $P_0(x)=\delta(x)$. Substituting the assumed forms of $q(x,t)$ and $W(x,t)$ in \eqref{1.9} and \eqref{1.10} into \eqref{1..6} and \eqref{1..7}, respectively, in Laplace space the iteration relation of $\widehat{T}_m(s)$ and the relation between $\widehat{R}_m(s)$ and $\widehat{T}_m(s)$ are given as
%
\begin{widetext}
\begin{equation}\label{1.11}
\begin{split}
   \sqrt{\pi} 2^m m!\widehat{T}_m(s)=& \frac{1}{2}\sum_{k=0}^{m}\frac{2^k \sqrt{\pi} m! }{(m-k)!}\left((2 v_0)^{m-k}+(-2 v_0)^{m-k}\right)\mathscr{L}_{t}\left\{ a(t)^{k-m} \mathscr{L}^{-1}_{s}\left[(-1)^{m-k}\hat{\phi}^{(m-k)}(s)\widehat{T}_k(s)\right](t)\right\}(s)\\
 &+H_m(0),
\end{split}
\end{equation}
where $\hat{\phi}^{(m-k)}(s)=\frac{d^{m-k}}{ds^{m-k}} \phi(s)$, and
\begin{equation}\label{1.12}
\sqrt{\pi} 2^m m!\widehat{R}_m(s)= \frac{1}{2}\sum_{k=0}^{m}\frac{2^k \sqrt{\pi} m! }{(m-k)!}\left((2 v_0)^{m-k}+(-2 v_0)^{m-k}\right)\mathscr{L}_{t}\left\{ a(t)^{k-m} \mathscr{L}^{-1}_{s}\left[(-1)^{m-k}\hat{\psi}^{(m-k)}(s)\widehat{T}_k(s)\right](t)\right\}(s).
\end{equation}
\end{widetext}
The detailed derivations are shown in Appendix \ref{App_B}.

The long time asymptotic behaviors of the first two moments in both coordinates attract our interest. Firstly, we concentrate on the first two moments for the process in comoving coordinate. From \cite{Zaburdaev}, there exists
\begin{equation}\label{mmt}
	\langle \hat{x}^m(s)\rangle=(i)^m \left. \frac{\partial^m}{\partial k^m} \widehat{\widetilde{W}}(k,s)\right|_{k=0}.
\end{equation}
Further, the specific form of $W(x,t)$ can be obtained by combining \eqref{1.10} with \eqref{a1}, which is
\begin{equation}\label{1.13}
  W(x,t)=\sum_{n=0}^{\infty}(-1)^n \frac{d^n}{d x^n} e^{-x^2} R_n(t).
\end{equation}
Applying the Fourier transform $x\to k$ and Laplace transform $t\to s$ on \eqref{1.13} lead to
\begin{equation}\label{1.14}
  \widehat{\widetilde{W}}(k,s)=\sum_{n=0}^{\infty} \sqrt{\pi}(-i k)^n e^{-\frac{k^2}{4}} \widehat{R}_n(s).
\end{equation}
The first two comoving moments of the process in Laplace space can be further represented as
\begin{equation}\label{firstmoment}
   \langle \hat{x}(s)\rangle =i \left.\frac{\partial}{\partial k} \widehat{\widetilde{W}}(k,s)\right|_{k=0}=\sqrt{\pi} \widehat{R}_1(s)
\end{equation}
and
\begin{equation}\label{msd}
\begin{split}
   \langle\hat{x}^2(s)\rangle & =- \frac{\partial^2}{\partial k^2} \widehat{\widetilde{W}}(k,s)|_{k=0}\\
   &=\frac{\sqrt{\pi}}{2} \widehat{R}_0(s)+2\sqrt{\pi} \widehat{R}_2(s).
\end{split}
\end{equation}

In order to calculate the first two moments of the process in comoving coordinate, the values of $\widehat{R}_0(s), \widehat{R}_1(s)$, and $\widehat{R}_2(s)$ should be calculated first. Taking $m=0,1,2$ in \eqref{1.11} and \eqref{1.12} results in
\begin{equation}\label{1.16}
\begin{split}
   &\widehat{T}_0(s)=\frac{1}{\sqrt{\pi}\big(1-\hat{\phi}(s)\big)}, \\
     & \widehat{R}_0(s)=\frac{1}{s\sqrt{\pi}},\\
     &\widehat{T}_1(s)=\widehat{R}_1(s)=0.
\end{split}
\end{equation}
Additionally,
\begin{equation}\label{1.17}
\begin{split}
   & \sqrt{\pi}2^3\left(1-\hat{\phi}(s)\right) \widehat{T}_2(s)\\
    & =\sqrt{\pi} (2 v_0)^2\mathscr{L}_{t}\left\{\frac{1}{a^2(t)}\mathscr{L}^{-1}_{s}\left[\hat{\phi}''(s)\widehat{T}_0(s)\right](t)\right\}(s)-2,
\end{split}
\end{equation}
\begin{equation}\label{1.18}
\begin{split}
   & \sqrt{\pi}2^3 \widehat{R}_2(s)- \sqrt{\pi}2^3\hat{\psi}(s)\widehat{T}_2(s)\\
   & =\sqrt{\pi} (2 v_0)^2\mathscr{L}_{t}\left\{\frac{1}{a^2(t)}\mathscr{L}^{-1}_{s}\left[\hat{\psi}''(s)\widehat{T}_0(s)\right](t)\right\}(s).\\
\end{split}
\end{equation}
%
Moreover, the first moment in comoving coordinate $\langle x(t)\rangle=0$ since $\widehat{R}_1(s)=0$, which indicates the process in comoving coordinate is symmetric. It should be noted that the normalization of the PDF $W(x,t)$ can also be verified through Hermite polynomials expansion and the result of $R_0=1/\sqrt{\pi}$, the detailed calculations of which can be found in \cite{Pengbo,Pengbo1}.

Next we change the results from comoving coordinate to the physical coordinate. In fact from \eqref{1.5}, the following relations of the first two moments in different coordinates can be easily obtained,
%
\begin{equation}\label{1..21}
  \langle y(t)\rangle=a(t)\langle x(t)\rangle
\end{equation}
and
\begin{equation}\label{1.21}
  \langle y^2(t)\rangle=a^2(t)\langle x^2(t)\rangle.
\end{equation}
Obviously the average moment $\langle y(t)\rangle=0$ since $\langle x(t)\rangle=0$, which means the expansion or contraction of medium does not change the symmetry of the process. It can be easily noticed that the PDF $P(y,t)$ is also normalized from the normalization of $W(x,t)$.

In the following, by considering some representative walking time PDFs $\phi(\tau)$ and scale factors $a(t)$, we discuss the long-time asymptotic behaviors of the MSDs in both comoving and physical coordinates.
%

\section{Dynamics of L\'evy walk in non-static medium} \label{sec 4}

\subsection{Exponentially distributed walking time}

In this subsection the walking time PDF is assumed to be exponential distribution, i.e., $\phi(\tau)=\lambda e^{-\lambda\tau}$, with $\lambda>0$. For different scale factors $a(t)$, we are going to analyze the asymptotic behavior of the statistical properties of the process in comoving and physical coordinates.

\subsubsection{Exponential scale factor}

We focus on the case that the non-static medium is described by the exponential scale factor
\begin{equation}\label{1.19}
  a(t)=\exp(H t)
\end{equation}
with the ``Hubble constant'' $H=a'(t)/a(t)$ \cite{cosmological, Introducing}. The case $H>0$ corresponds to an expanding medium while the case $H<0$ represents a contracting one.

\begin{figure}[htbp]
\centering
\includegraphics[scale=0.28]{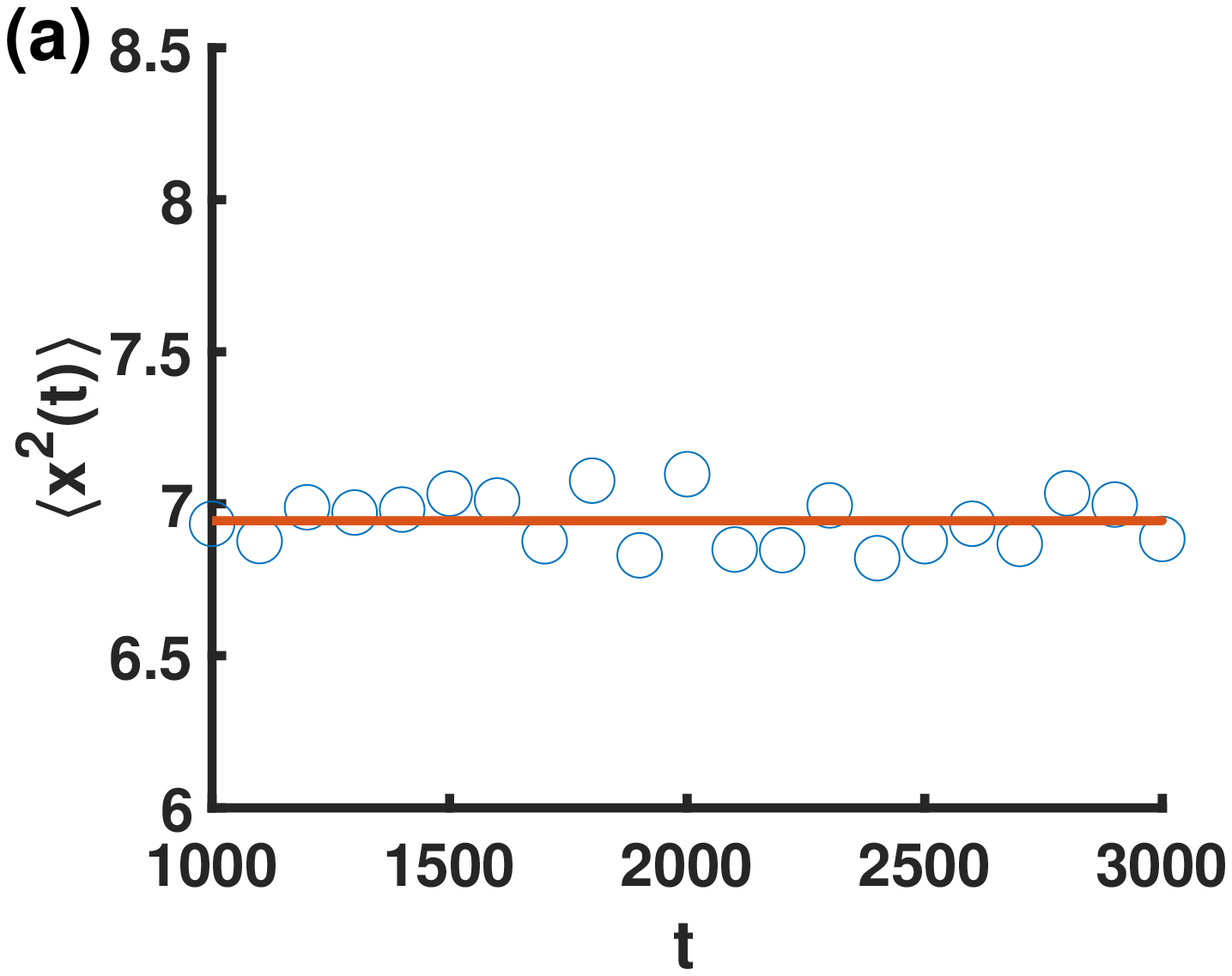}
\includegraphics[scale=0.28]{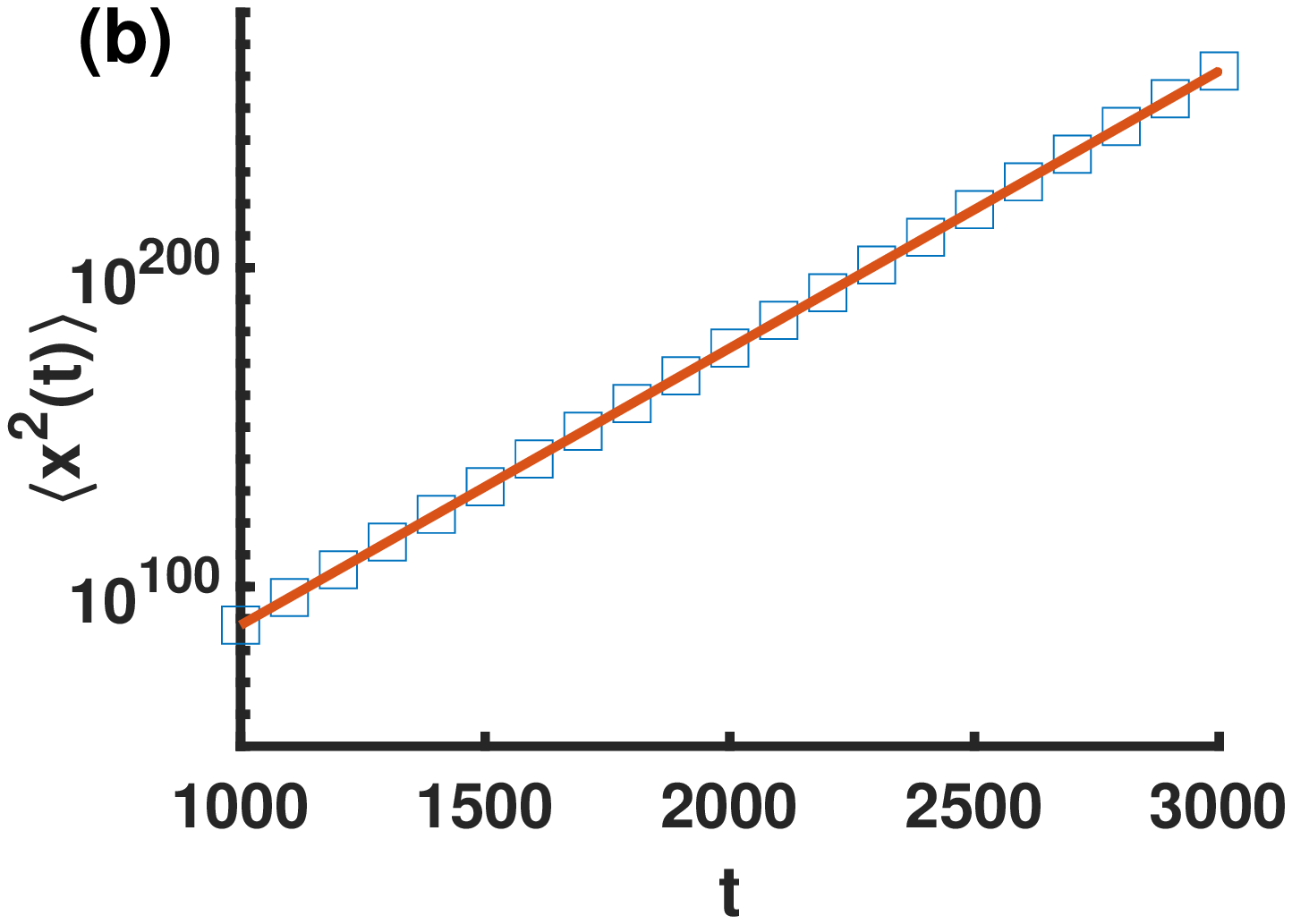}
\includegraphics[scale=0.28]{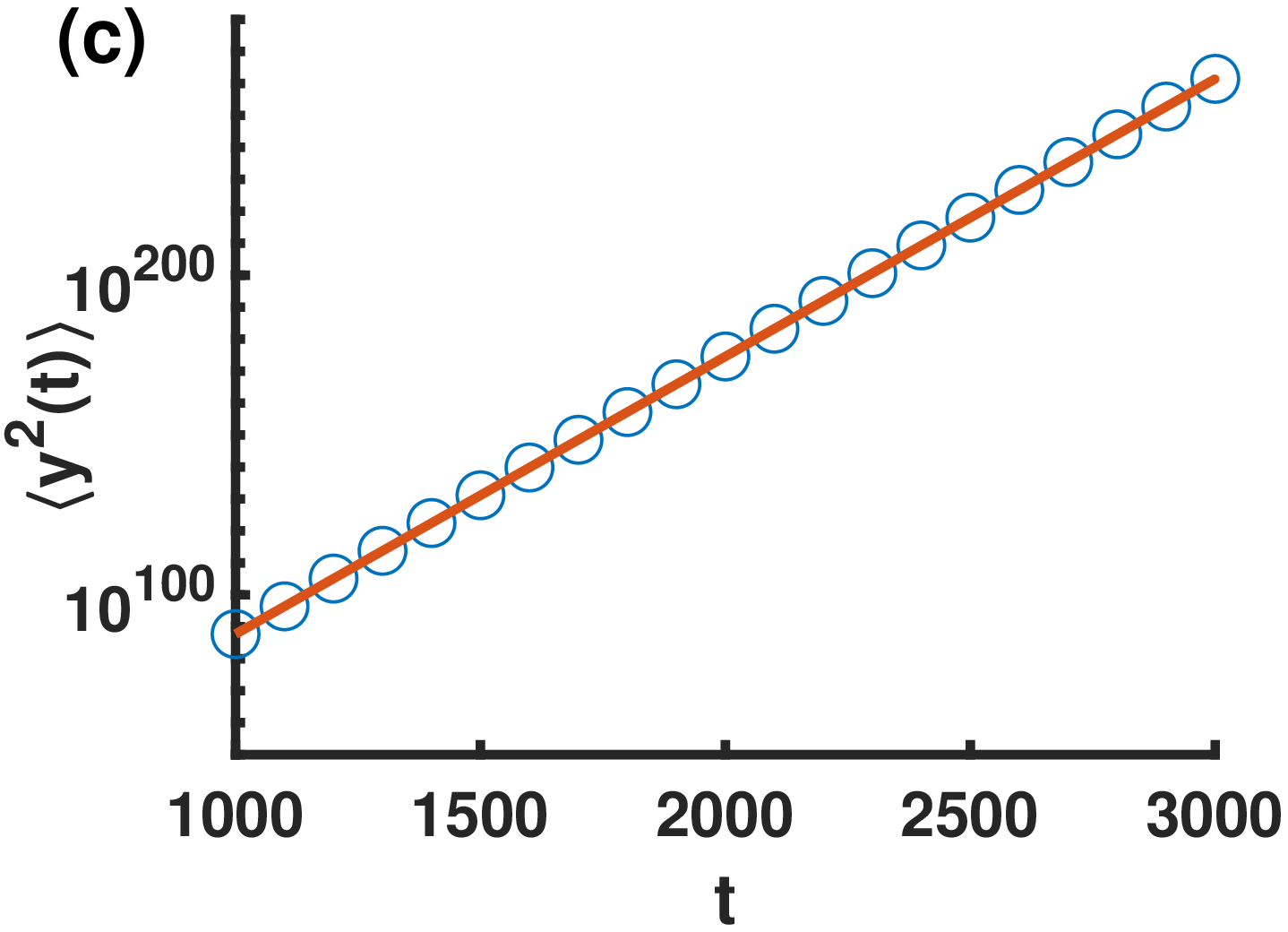}
\includegraphics[scale=0.28]{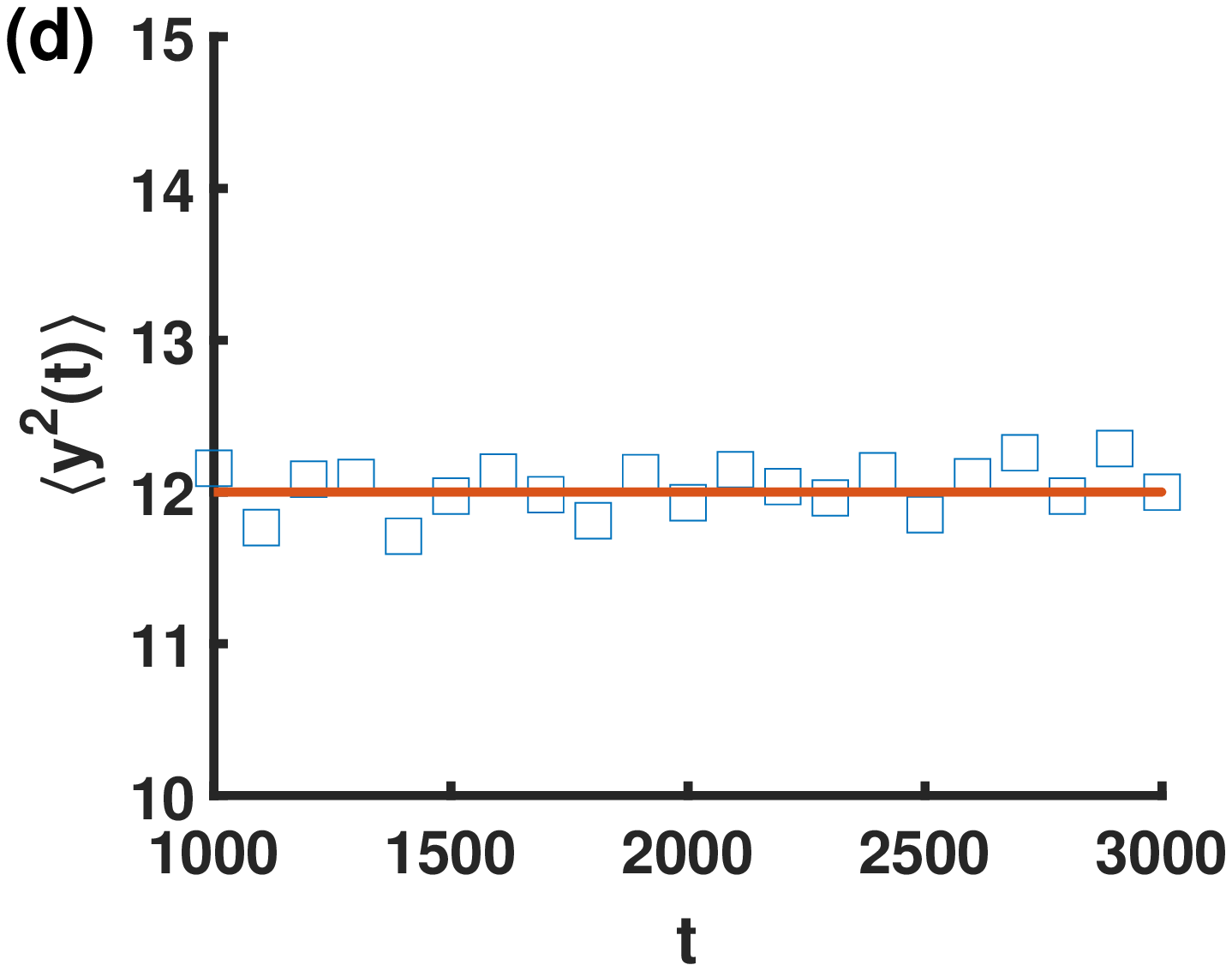}
\caption{Numerical simulations of the MSDs of L\'{e}vy walk in uniform expansion medium with exponential scale factor by sampling over $10^4$ realizations. The walking time PDF of L\'{e}vy walk behaves as exponential distribution $\phi(\tau)=\lambda e^{-\lambda\tau}$ with $\lambda=1$. The parameters are $v_0=1$ and $x_0=y_0=0$. For (a) and (c), we take $H=0.1$; for (b) and (d), $H=-0.1$. The solid lines in (a), (b) and (c), (d) respectively are the theoretical results shown in \eqref{1.20} and \eqref{1.22}.}
\label{expexp}
\end{figure}

Combining \eqref{msd} with  \eqref{1.16}, \eqref{1.17}, as well as \eqref{1.18}, the MSDs in comoving coordinate can be got
\begin{equation}\label{1.20}
\langle x^2(t)\rangle\sim
\left\{
\begin{split}
  &\frac{\lambda v_0^2}{H (2 H+\lambda)^2}, \,\quad\quad\quad \mbox{if $H>0$},  \\
  & \frac{(2 H-\lambda)v_0^2}{H \lambda^2}e^{-2 H t}, \quad \mbox{if $H<0$}.
\end{split}
\right.
\end{equation}
The results in \eqref{1.20} are supported by Fig. \ref{expexp}(a) and (b). It can be concluded from \eqref{1.20} that for $H>0$ the comoving variance tends to be a constant for sufficient long time which relies on the value of $H$ and $\lambda$ as well as the velocity $v_0$ of L\'{e}vy walks. In contrast, the MSD in comoving coordinate displays exponential growth for $H<0$. The similar conclusions for subdiffusion CTRW in non-static medium can be found in \cite{F}.

Now we turn to analyzing MSDs in physical coordinate. Combining \eqref{1.20} with the relation \eqref{1.21}, we have
\begin{equation}\label{1.22}
\langle y^2(t)\rangle\sim
\left\{
\begin{split}
  & \frac{\lambda v_0^2}{H (2 H+\lambda)^2}e^{2 H t}, \quad \mbox{if $H>0$},  \\
  & \frac{(2 H-\lambda)v_0^2}{H \lambda^2}, \quad \quad \quad ~ \mbox{if $H<0$},
\end{split}
\right.
\end{equation}
which have been verified by Fig. \ref{expexp}(c) and (d).
For $H>0$, the variance in physical space exponentially increases with respect to the time $t$. However, the MSD keeps a constant for sufficiently long time for $H<0$, which implies a stationary propagator function in physical space can be reached for long time limit. Compared with the MSD of L\'{e}vy walk in static medium which behaves as $\langle y^2(t)\rangle \sim t$, one can conclude that the dominant term of the diffusive processes is the displacement induced by exponential scale factor of the medium.

\begin{figure}[htbp]
\centering
\includegraphics[scale=0.19]{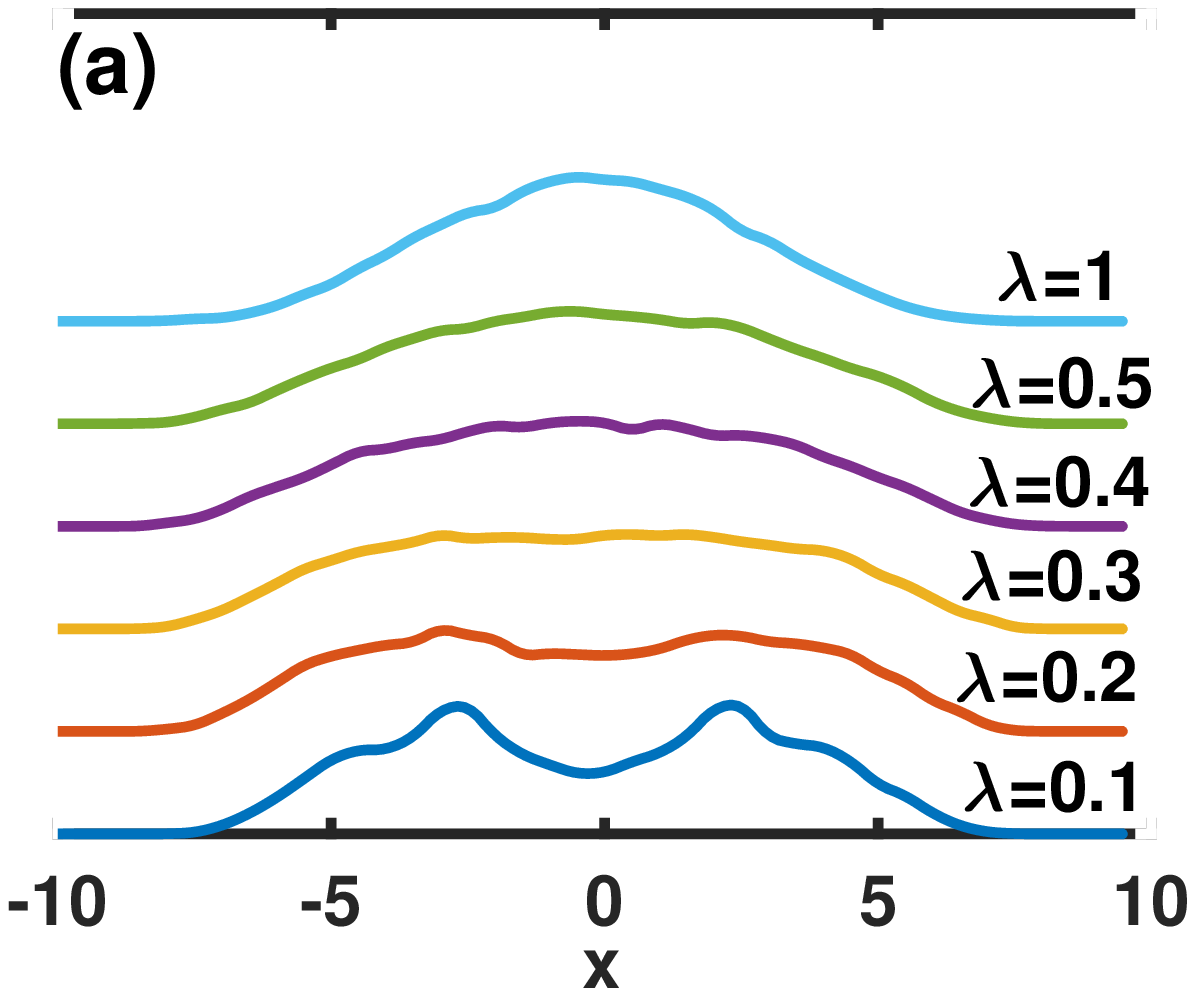}
\includegraphics[scale=0.19]{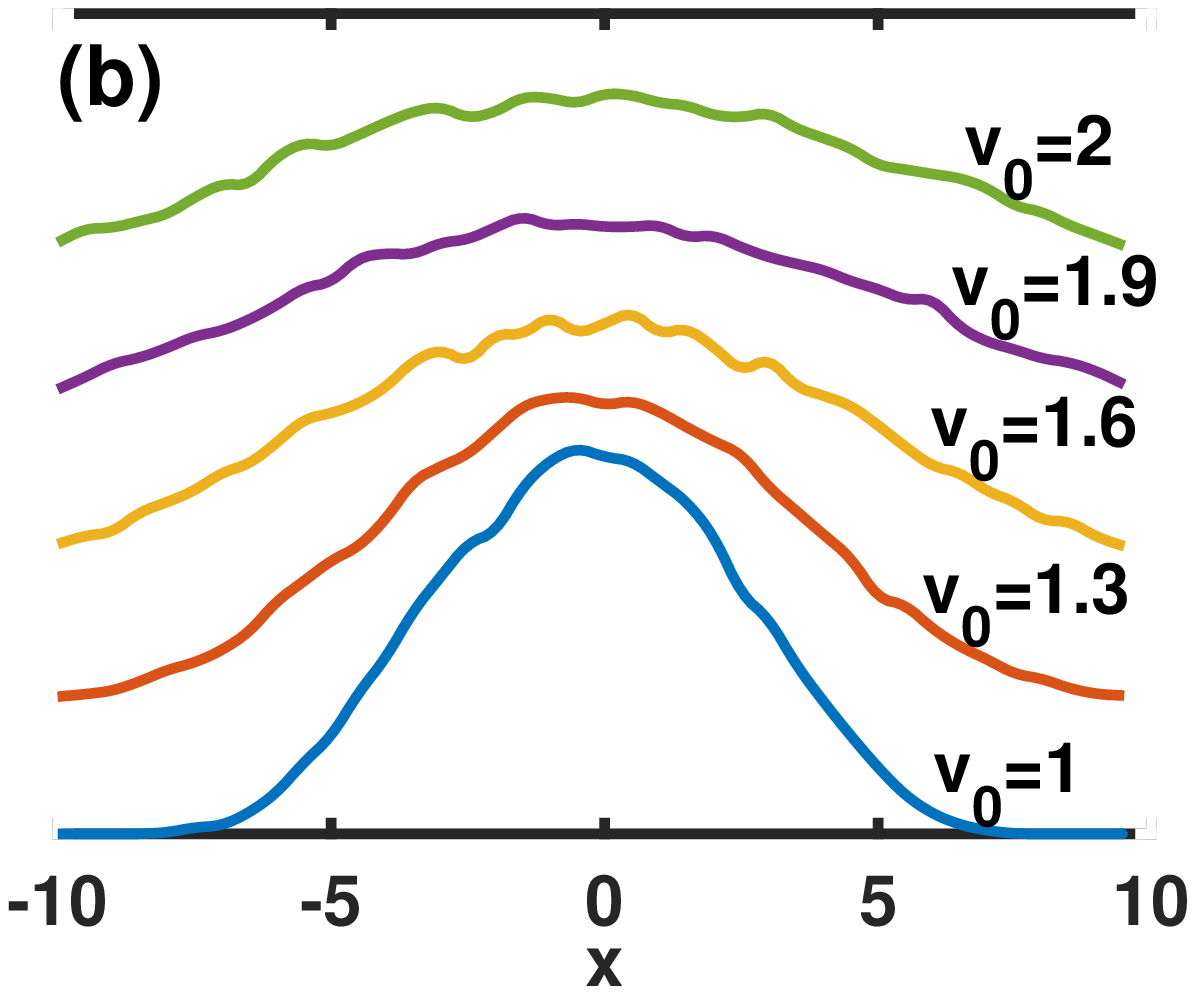}
\includegraphics[scale=0.19]{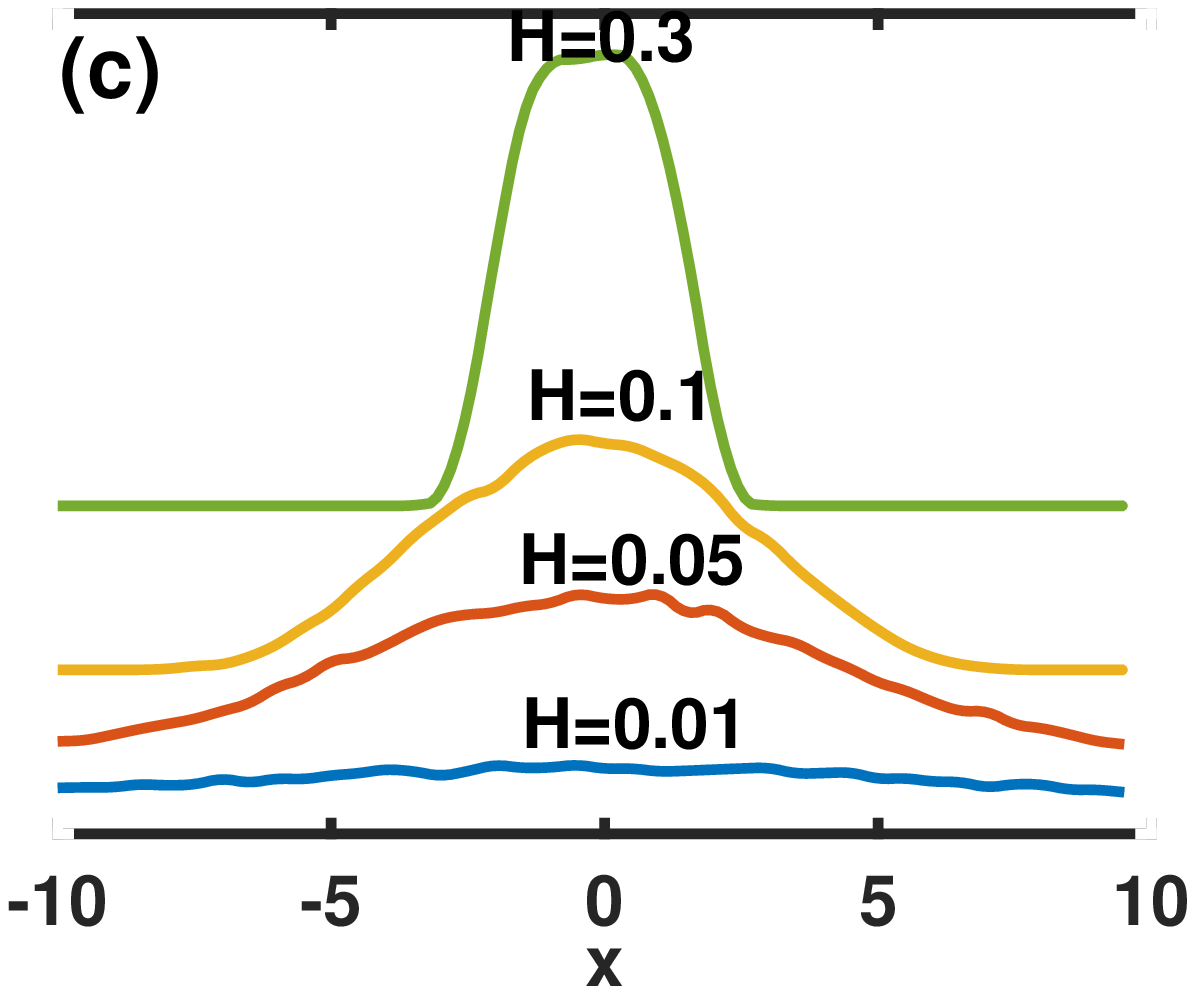}
\caption{Stationary PDFs of L\'{e}vy walk with $\phi(\tau)=\lambda e^{-\lambda\tau}$ and exponentially expanding scale factor in comoving coordinates by sampling over $2\times10^4$ realizations. Here we assume $x_0=0$. For panel (a), we take $H=0.1$, $v_0=1$; for panel (b), we use $H=0.1$, $\lambda=1$; for panel (c),  $\lambda=1$ and $v_0=1$.}
\label{pstm}
\end{figure}
\begin{figure}[htbp]
\centering
\includegraphics[scale=0.19]{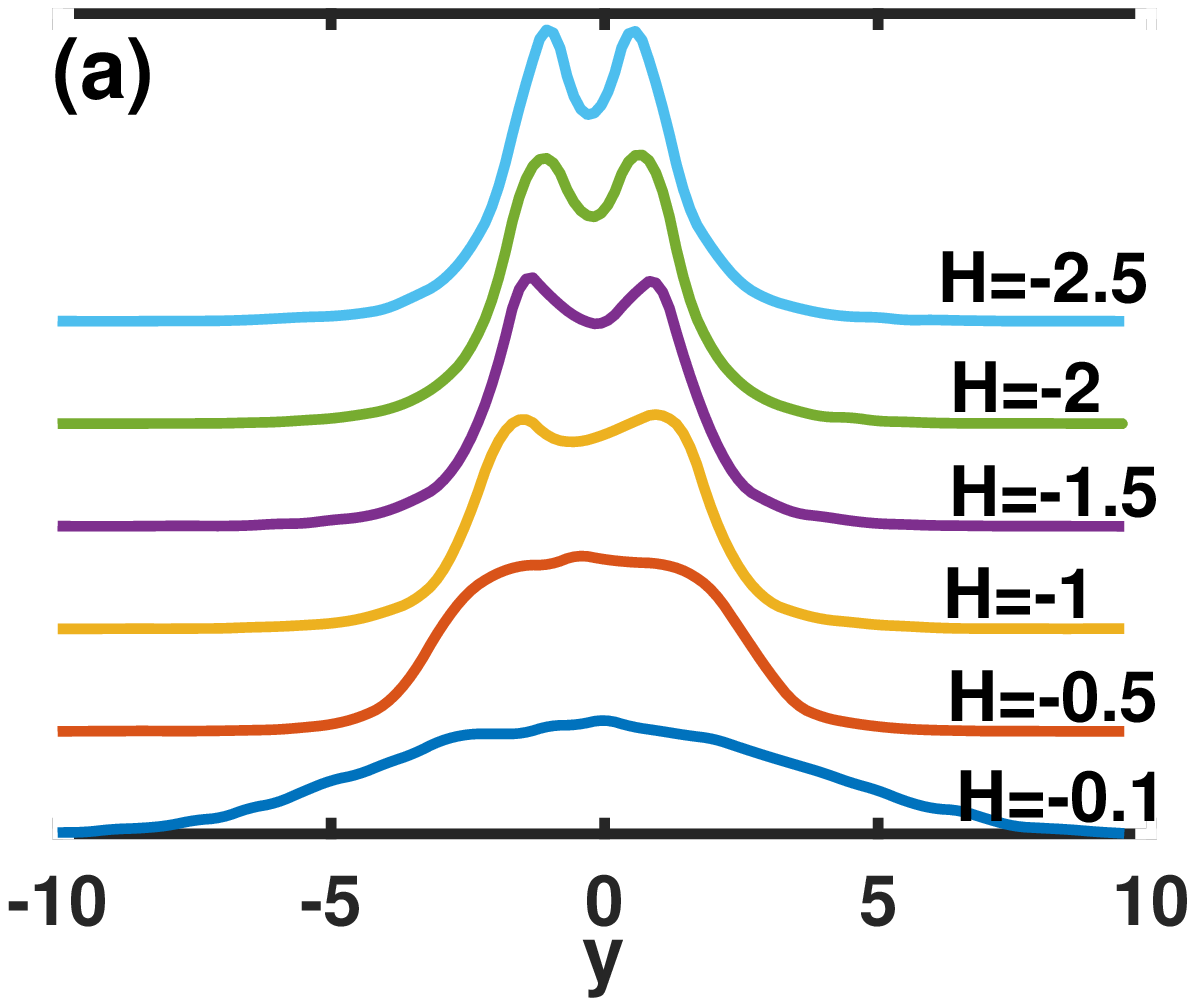}
\includegraphics[scale=0.19]{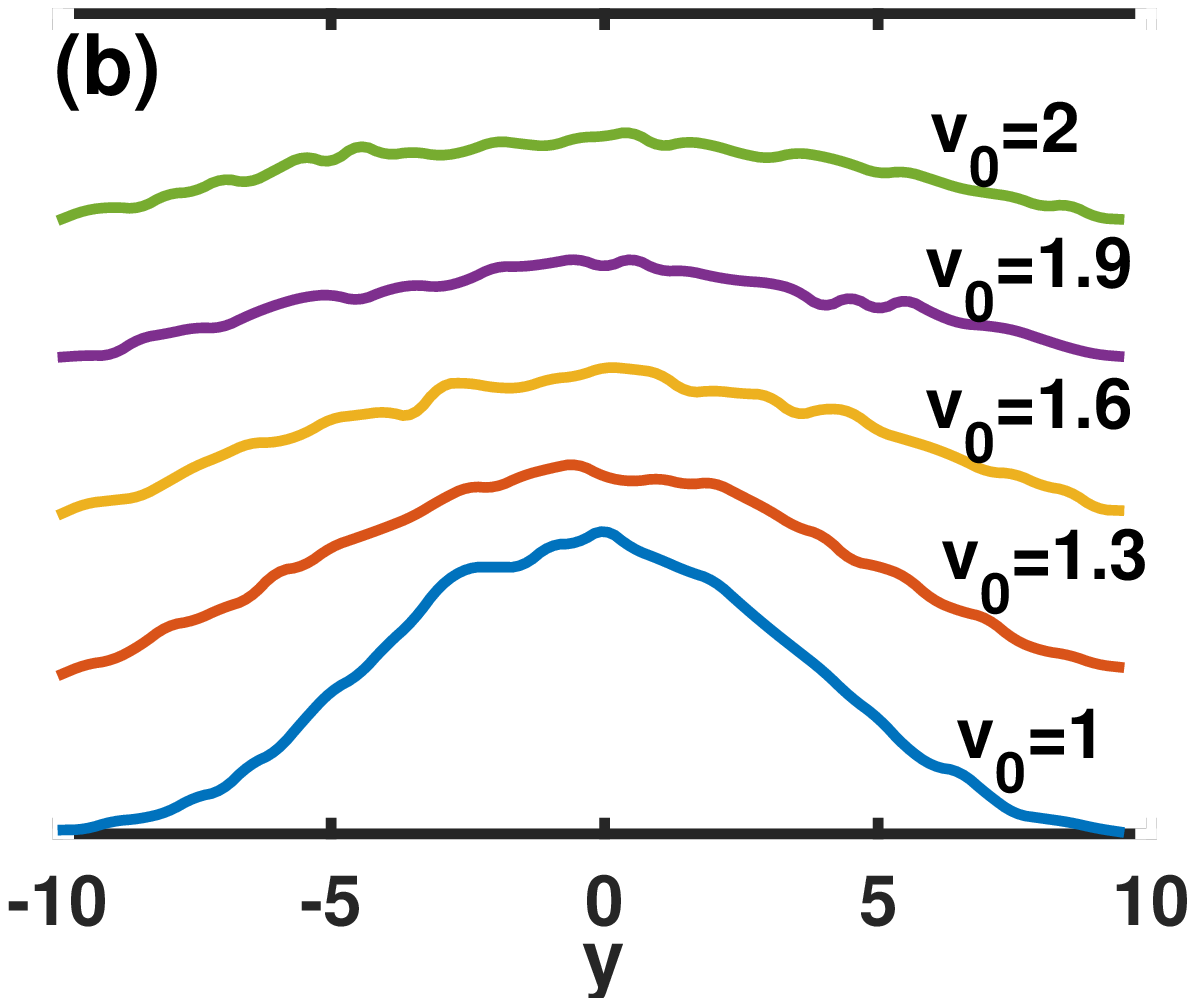}
\includegraphics[scale=0.19]{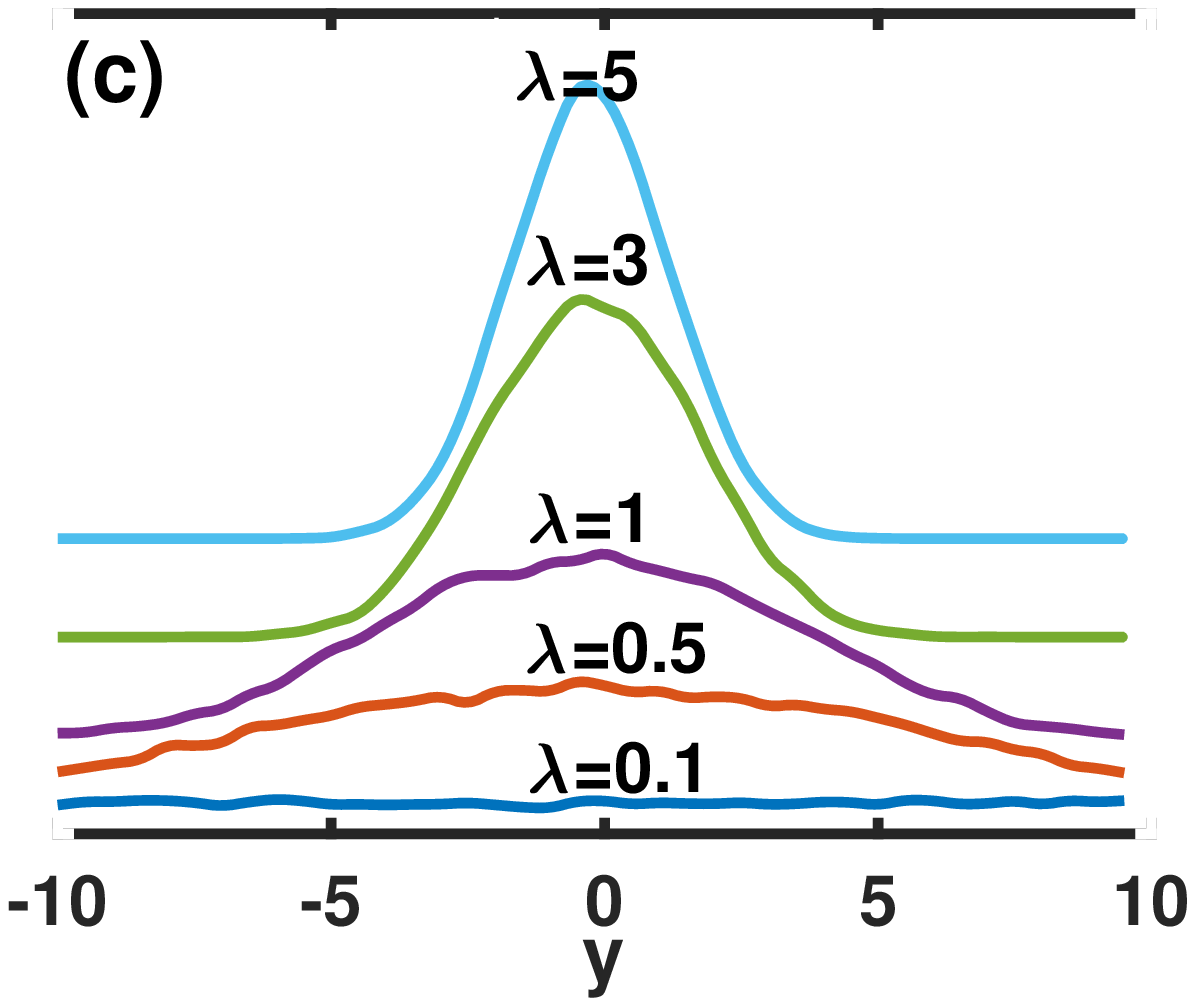}
\caption{Stationary PDFs of L\'{e}vy walk with $\phi(\tau)=\lambda e^{-\lambda\tau}$ and exponentially contracting scale factor in physical coordinates by sampling over $2\times10^4$ realizations. Here we assume $y_0=0$. For panel (a), we take $v_0=1$ and $\lambda=1$; for panel (b), we use $H=-0.1$ and $\lambda=1$; for panel (c),  $H=-0.1$ and $v_0=1$;.}
\label{pstym}
\end{figure}

Next we numerically simulate the stationary distribution $P^{st}(x)$ in comoving or physical space.  
For L\'{e}vy walk with $\phi(\tau)=\lambda e^{-\lambda\tau}$ in comoving coordinate and a positive $H$, the role of parameter $\lambda$ can be concluded from Fig. \ref{pstm}(a), specifically the increase of $\lambda$ changes $P^{st}(x)$ from bimodal distribution to unimodal one. While Fig. \ref{pstm}(b) indicates that the change of velocity $v_0$ does not make the stationary distribution become from bimodal to unimodal or vice versa; increasing $v_0$ can only flatten $P^{st}(x)$. Similarly, it can be concluded from Fig. \ref{pstm}(c) that the variation of Hubble constant $H>0$ can neither change the unimodality of stationary distribution, and the decrease of $H$ also plays the role of flattening $P^{st}(x)$. Interestingly, the stationary distribution in physically coordinated $P^{st}(y)$ for L\'evy walk process with $H<0$ has different properties. Specifically, according to Fig. \ref{pstym}(a) we conclude that the increasing absolute value of Hubble constant $|H|$ changes $P^{st}(y)$ from unimodal distribution to bimodal one. The increase of $v_0$ can only flatten the stationary distribution concluded from Fig. \ref{pstym}(b), and the unimodal or bimodal property cannot be affected by variation of $v_0$. The role of $\lambda$ turns out to be completely different from the one in comoving coordinate, specifically from Fig. \ref{pstym}(c) the stationary distribution is flattened as the parameter $\lambda$ decreases whereas the unimodal or bimodal property keeps the same with the change of $\lambda$.

Another quantity to characterize the property of PDF for stochastic process is kurtosis $K$, which measures the tails of the PDF. For a symmetric process with zero spatial average, the kurtosis can be defined as \cite{Pengbo1,Tian2021}
\begin{equation}\label{K}
	K=\frac{\langle y^4(t)\rangle}{\langle y^2(t)\rangle^2}.
\end{equation}
It is usually to choose $3$ as a benchmark obtained from normal distribution, and then make a comparison. 
Specifically for $K>3$ (or $K<3$), we call the corresponding PDF to be leptokurtic (or platykurtic). In this part, we only consider the kurtosis for the stationary density in physical coordinate, therefore we choose $H<0$. According to the definition of kurtosis, the fourth moment $\langle y^4(t)\rangle$ is required additionally. Similarly we first consider the corresponding statistical quantity in comoving coordinate $\langle x^4(t)\rangle$, and then convert it to the one in physical coordinate. It can be obtained from \eqref{mmt} and \eqref{1.14} that
\begin{equation}\label{fourthmomentxs}
   \langle \hat{x}^4(s)\rangle \!=\frac{3}{4}\sqrt{\pi} \widehat{R}_0(s)+6\sqrt{\pi} \widehat{R}_2(s)+24\sqrt{\pi} \widehat{R}_4(s),
\end{equation}
in which $\widehat{R}_4(s)$ can be explicitly calculated through \eqref{1.11} and \eqref{1.12}. Then by applying inverse Laplace transform on \eqref{fourthmomentxs}, there exists
\begin{equation}\label{fourthmomentx}
    \langle x^4(t)\rangle \!\sim\!\frac{3e^{-4 H t}\!(32 H^4\!-\!40 H^3 \lambda \!+\!16 H^2\lambda^2\!-\!6 H \lambda^3\!+\!\lambda^4)\! v_0^4 }{H^2 (2 H-\lambda)^2 \lambda^4}.
\end{equation}
Finally according to the relation of \eqref{1.5}, the asymptotic form of fourth moment in physical space is
\begin{equation}\label{fourthmomenty}
\begin{split}
   \langle y^4(t)\rangle & = a^4(t) \langle x^4(t)\rangle\\
     & \sim\frac{3(32 H^4-40 H^3 \lambda+16 H^2 \lambda^2-6 H \lambda^3+\lambda^4) v_0^4 }{H^2 (2 H-\lambda)^2 \lambda^4}.
\end{split}
\end{equation}
Combining with the MSD in \eqref{1.22} with $H<0$, for sufficiently long time there exists
\begin{equation}\label{Kvalue}
  K\sim\frac{3(32 H^4-40 H^3 \lambda+16 H^2 \lambda^2-6 H \lambda^3+\lambda^4)}{(2 H-\lambda)^4},
\end{equation}
which is verified by Fig. \ref{K}. Therefore, the asymptotic behavior of kurtosis is determined by Hubble constant and the inverse average of exponentially distributed walking time. At first for any fixed $H<0$ we consider $K$ as a function of $\lambda$. As shown in Fig. \ref{K}(a), 
$\lim_{\lambda\to 0} K=6$ indicating that the corresponding stationary density is leptokurtic for small $\lambda$. Then kurtosis decreases as $\lambda$ increases until $K$ reaches the minimum $K_{\min}=27/4-3\sqrt{2}\approx 2.51$ at $\lambda=-2(1+\sqrt{2})H$, and the minimum has no relevance to $H$, so that the stationary distribution enters the platykurtic category. With the further increase of $\lambda$, $K$ increases to finally reach the limit $3$ since $\lim_{\lambda\to\infty} K=3$. On the other hand, for fixed $\lambda>0$, we consider $K$ as a function of $H$, then the limit can be easily obtained, $\lim_{H\to 0}K=3$, which has been verified in Fig. \ref{K}(b). Further when $|H|$ increases, $K$ decreases to the same minimum $K_{\min}$ at $H=-\lambda(\sqrt{2}-1)/2$, indicating that the stationary density is platykurtic. Afterwards, $K$ keeps increasing to the limit $\lim_{H\to -\infty} K=6$, therefore the stationary density changes from platykurtic category to leptokurtosis one.
%

\begin{figure}[htbp]
\centering
\includegraphics[scale=0.28]{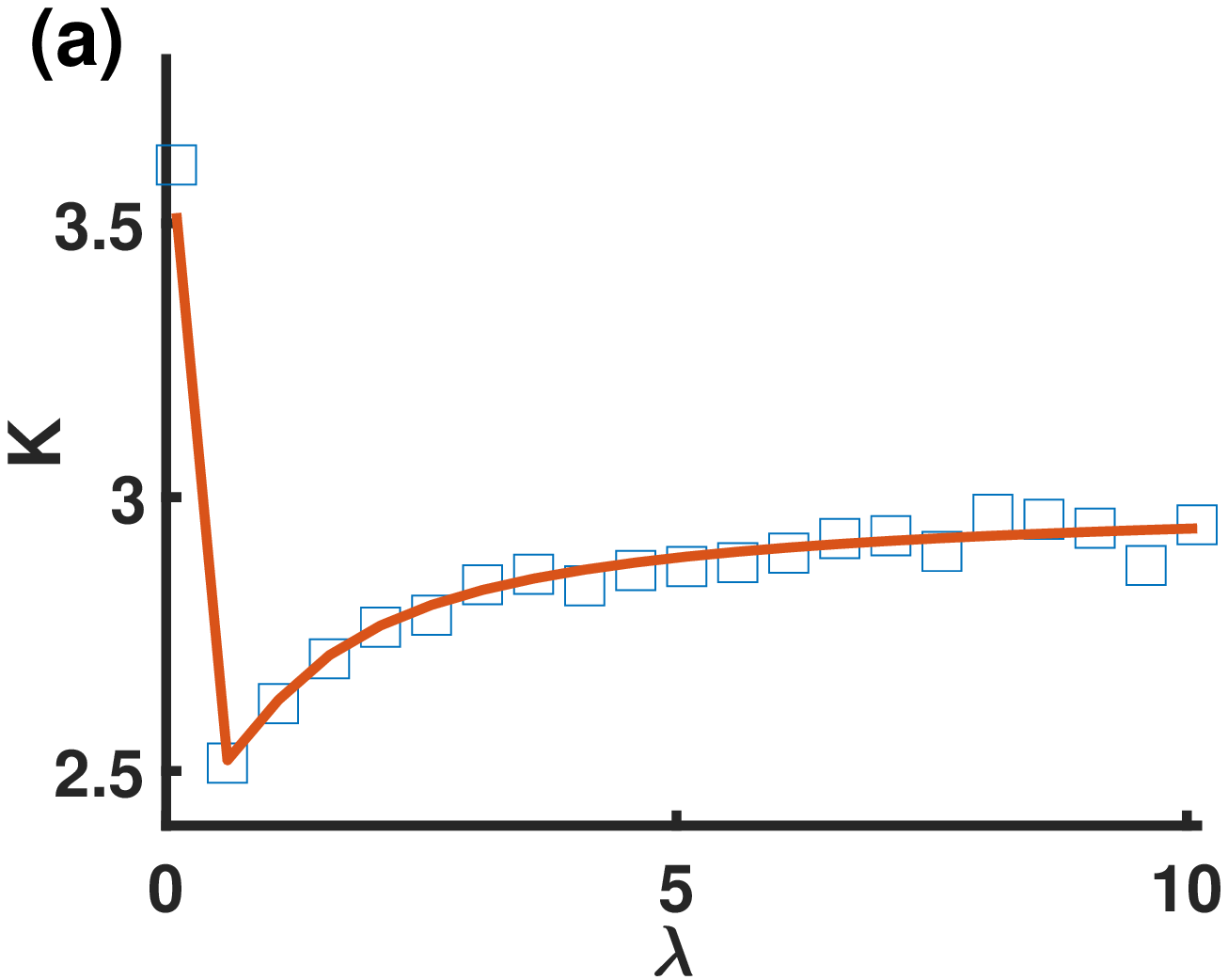}
\includegraphics[scale=0.28]{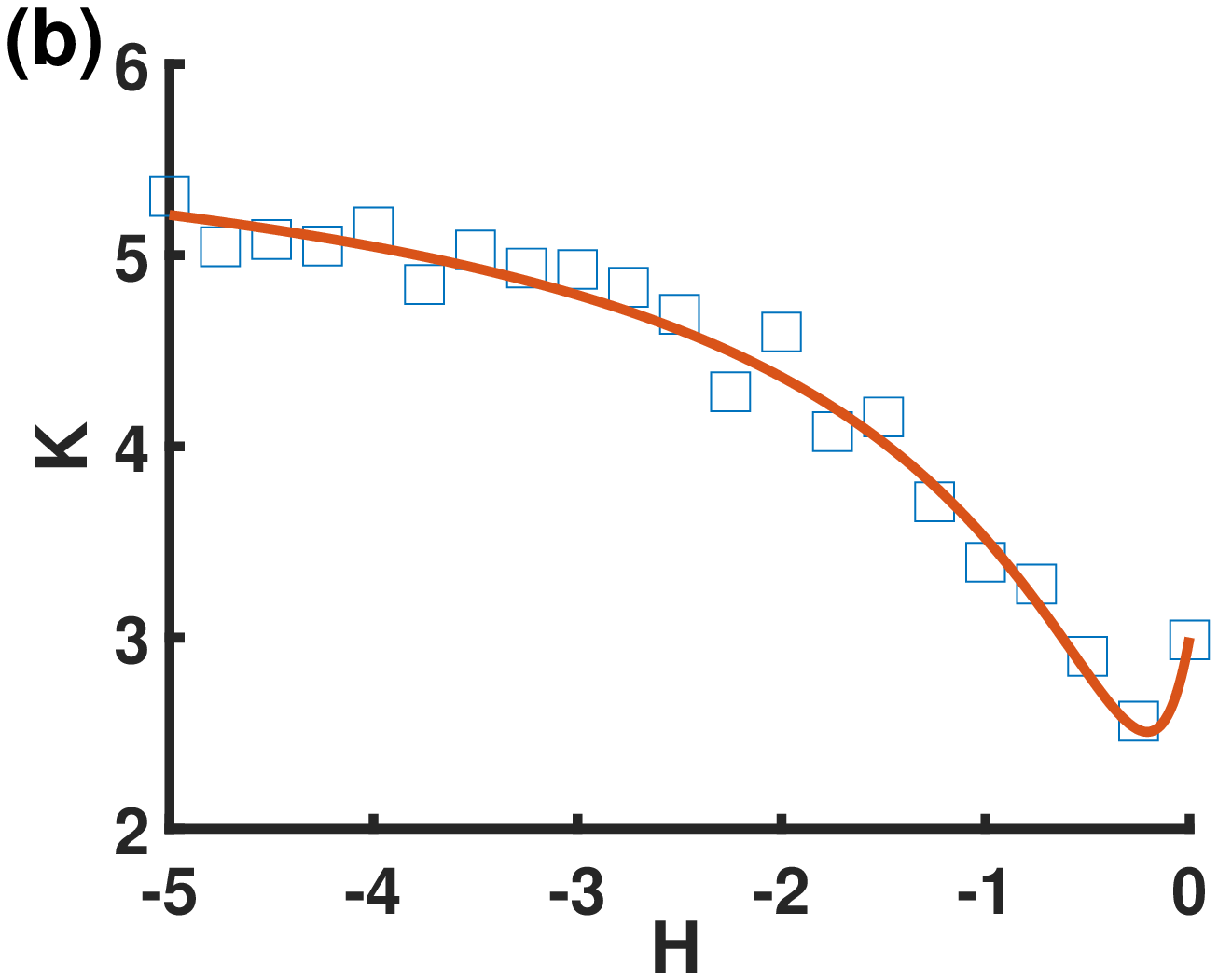}
\caption{Numerical simulations of kurtosis of L\'{e}vy walk in a non-static medium with exponential contraction by sampling over $3\times10^4$ realizations. The walking time PDF of L\'{e}vy walk behaves as exponential distribution $\phi(\tau)=\lambda e^{-\lambda\tau}$. The initial speed of L\'{e}vy walk is assumed to be $v_0=1$. For (a), we fix $H=-0.1$ and for (b) we take $\lambda=1$. The solid lines are the theoretical results \eqref{Kvalue}.}
\label{K}
\end{figure}

\subsubsection{Power-law scale factor}

Power-law scale factor also has many different types of applications in cosmology, which has the form
\begin{equation}\label{1.23}
  a(t)=\left(\frac{t+t_0}{t_0}\right)^\beta,
\end{equation}
where $t$ is the time elapsed after the initial time $t_0$ and the positive value of $\beta$ corresponds to an expanding medium while the negative value of $\beta$ describes a contracting one.

\begin{figure}[htbp]
\centering
\includegraphics[scale=0.28]{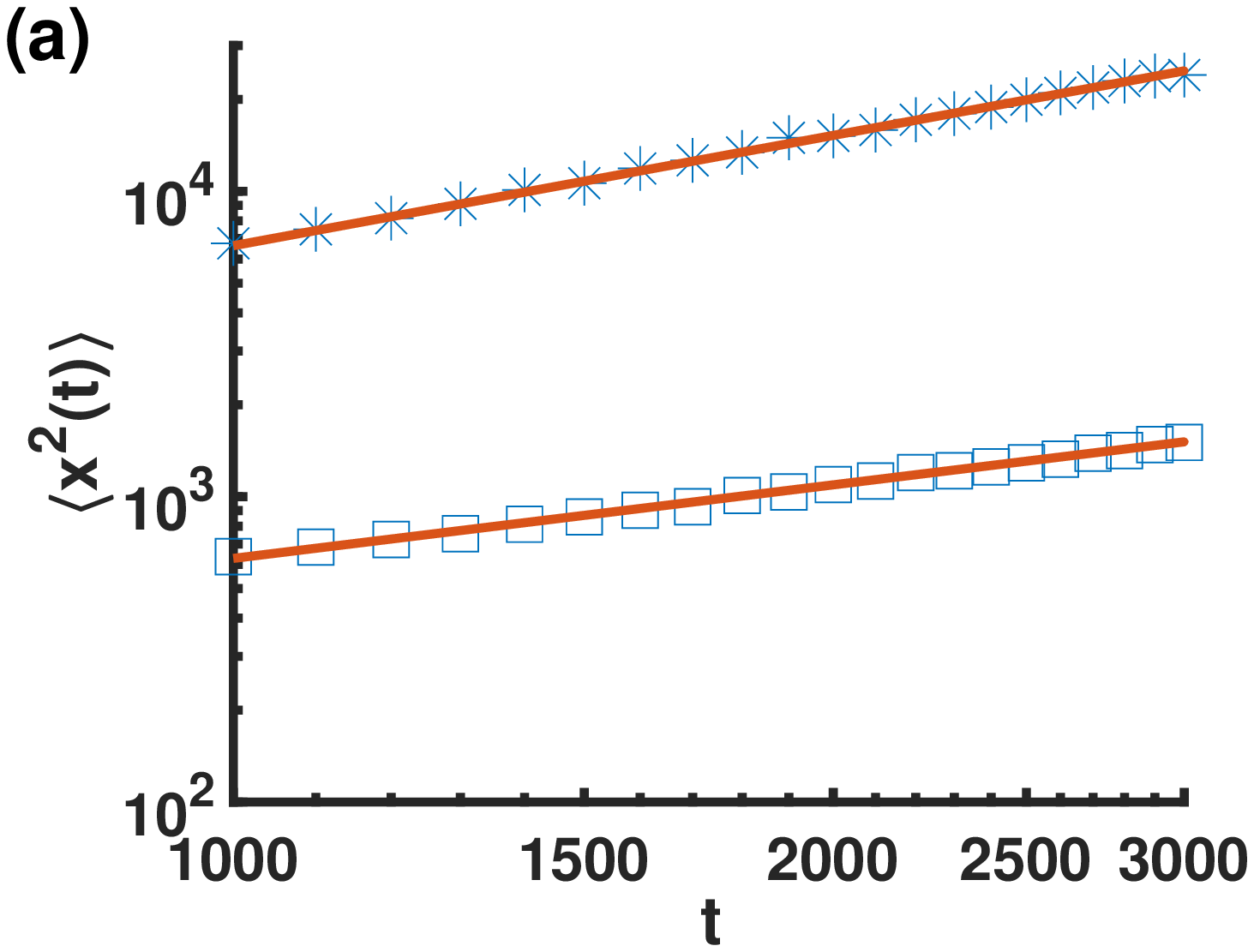}
\includegraphics[scale=0.28]{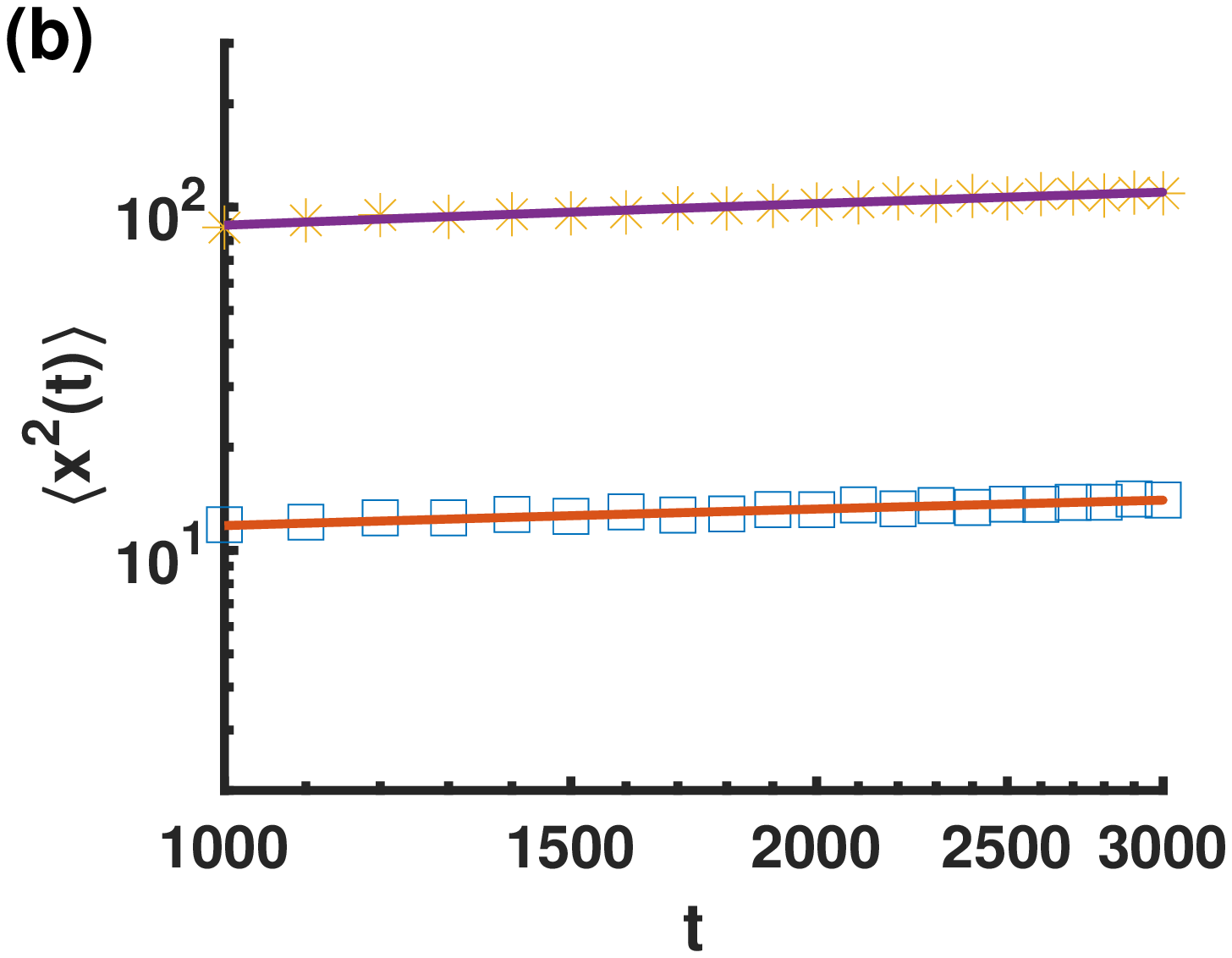}
\includegraphics[scale=0.28]{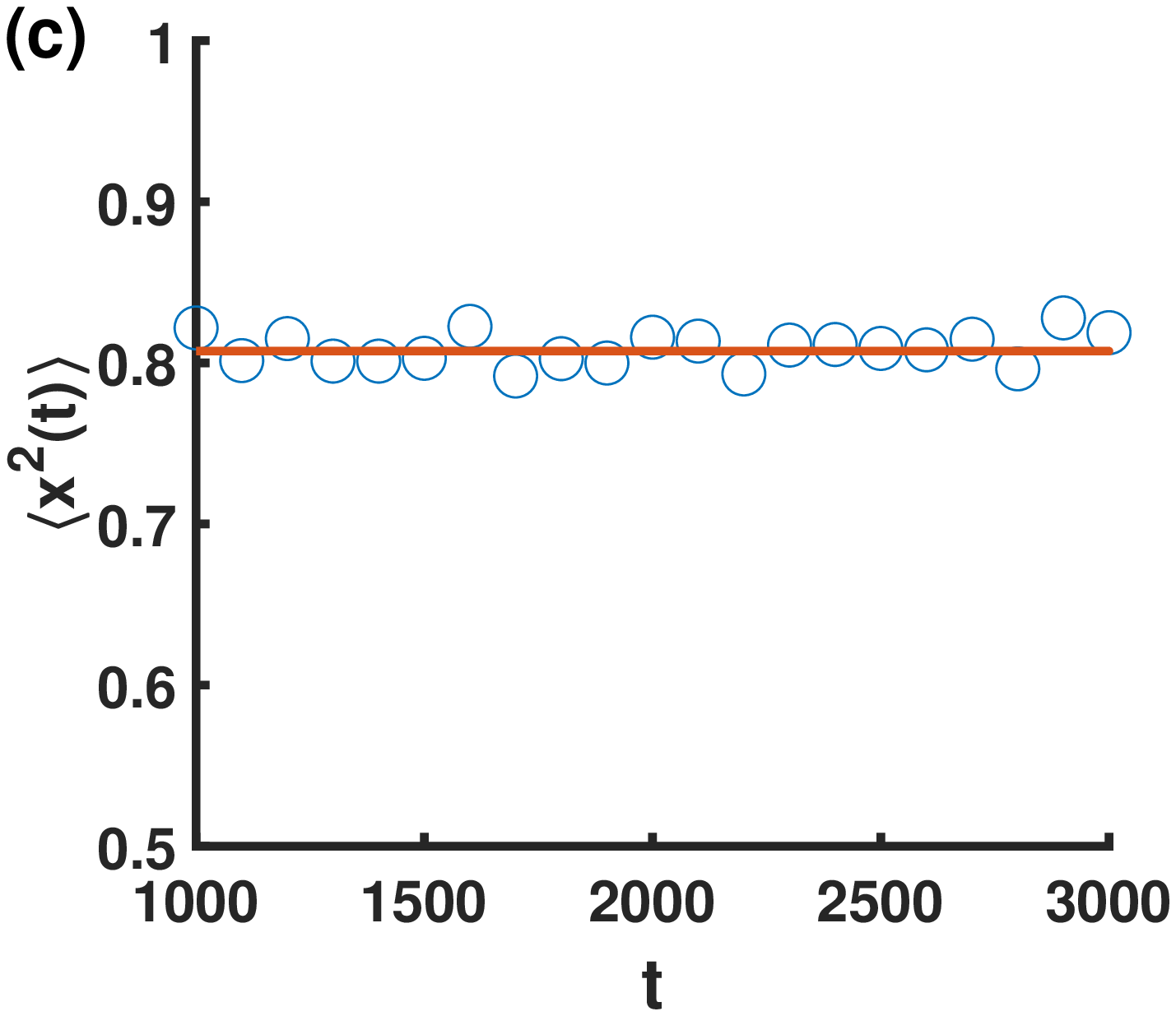}
\includegraphics[scale=0.28]{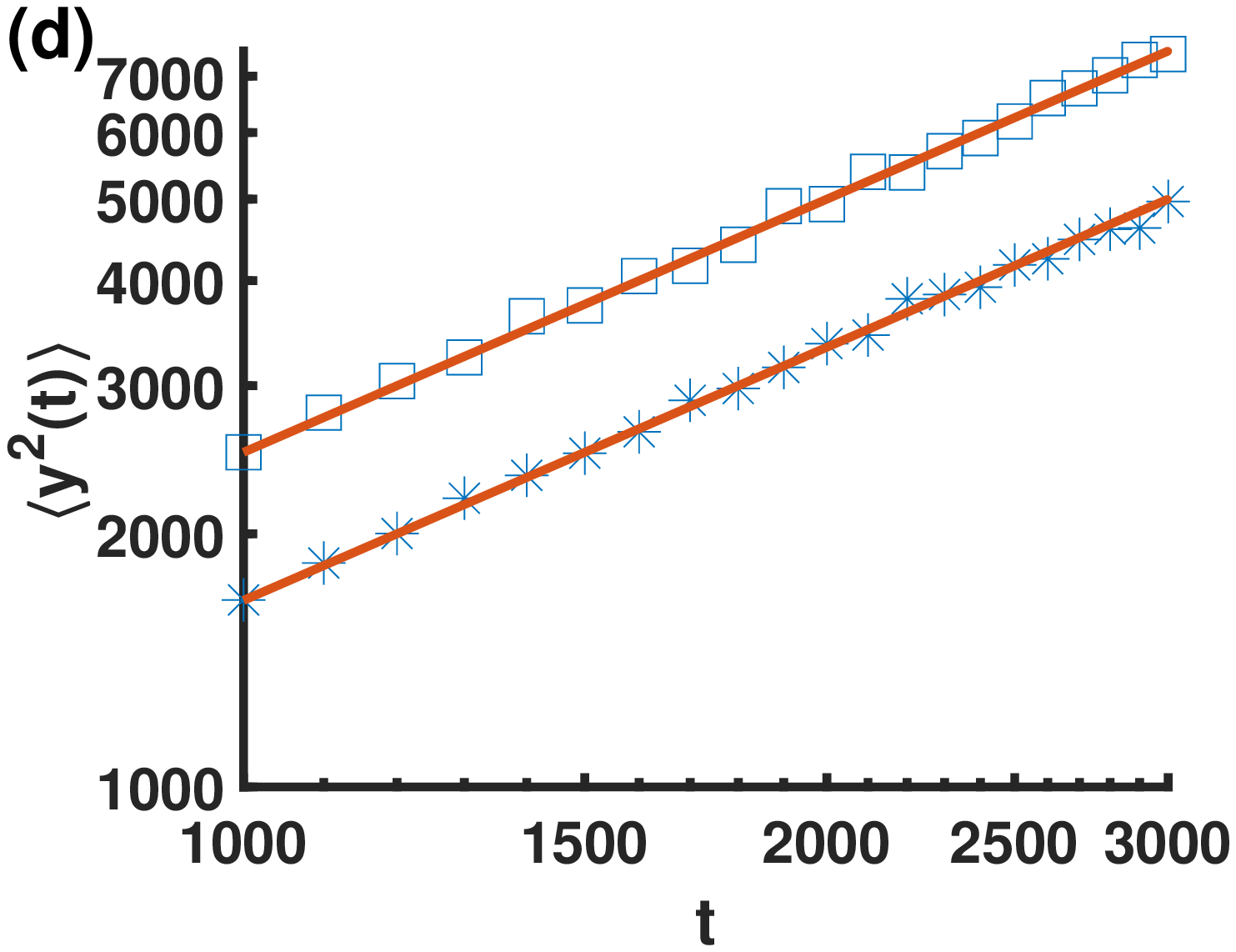}
\includegraphics[scale=0.28]{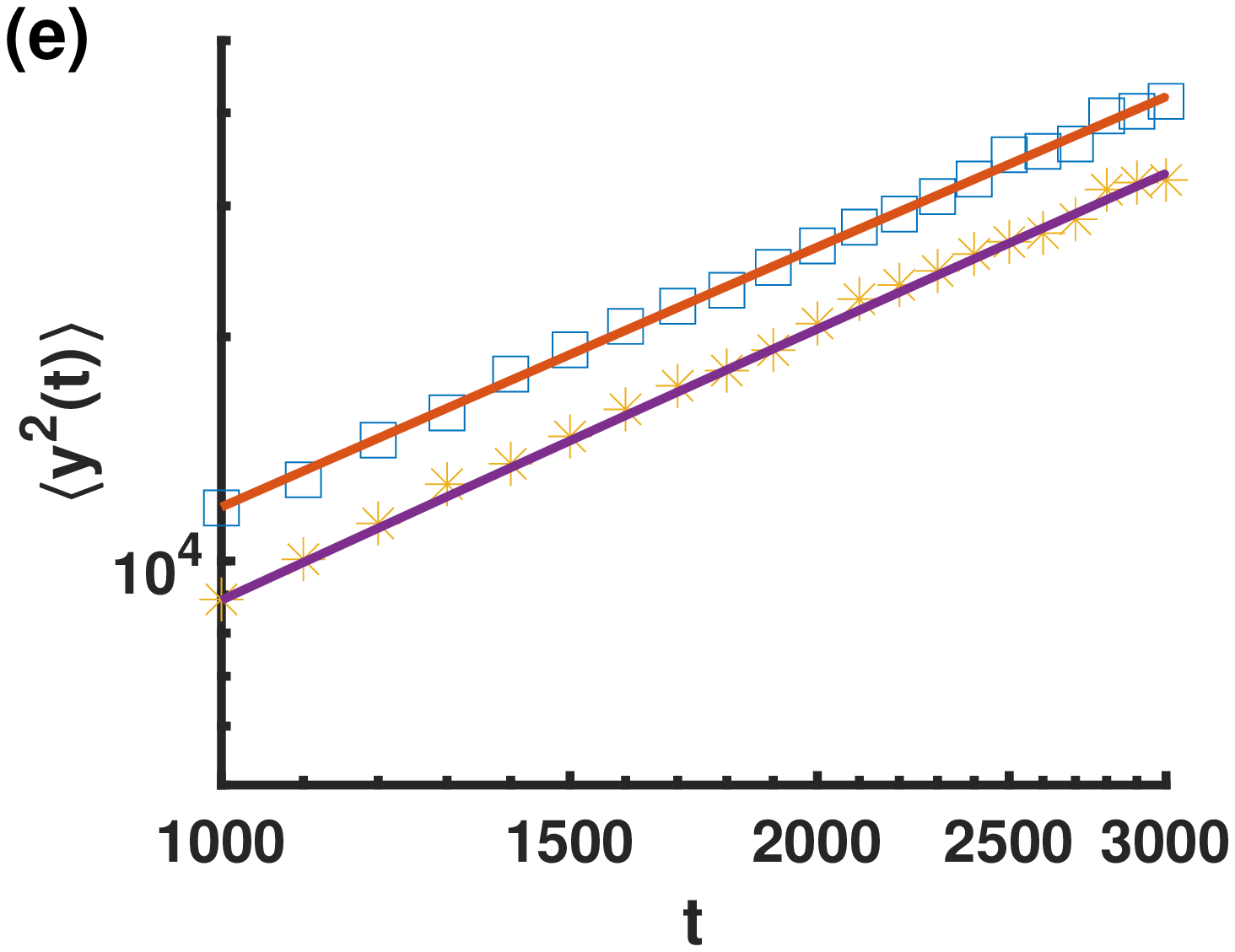}
\includegraphics[scale=0.28]{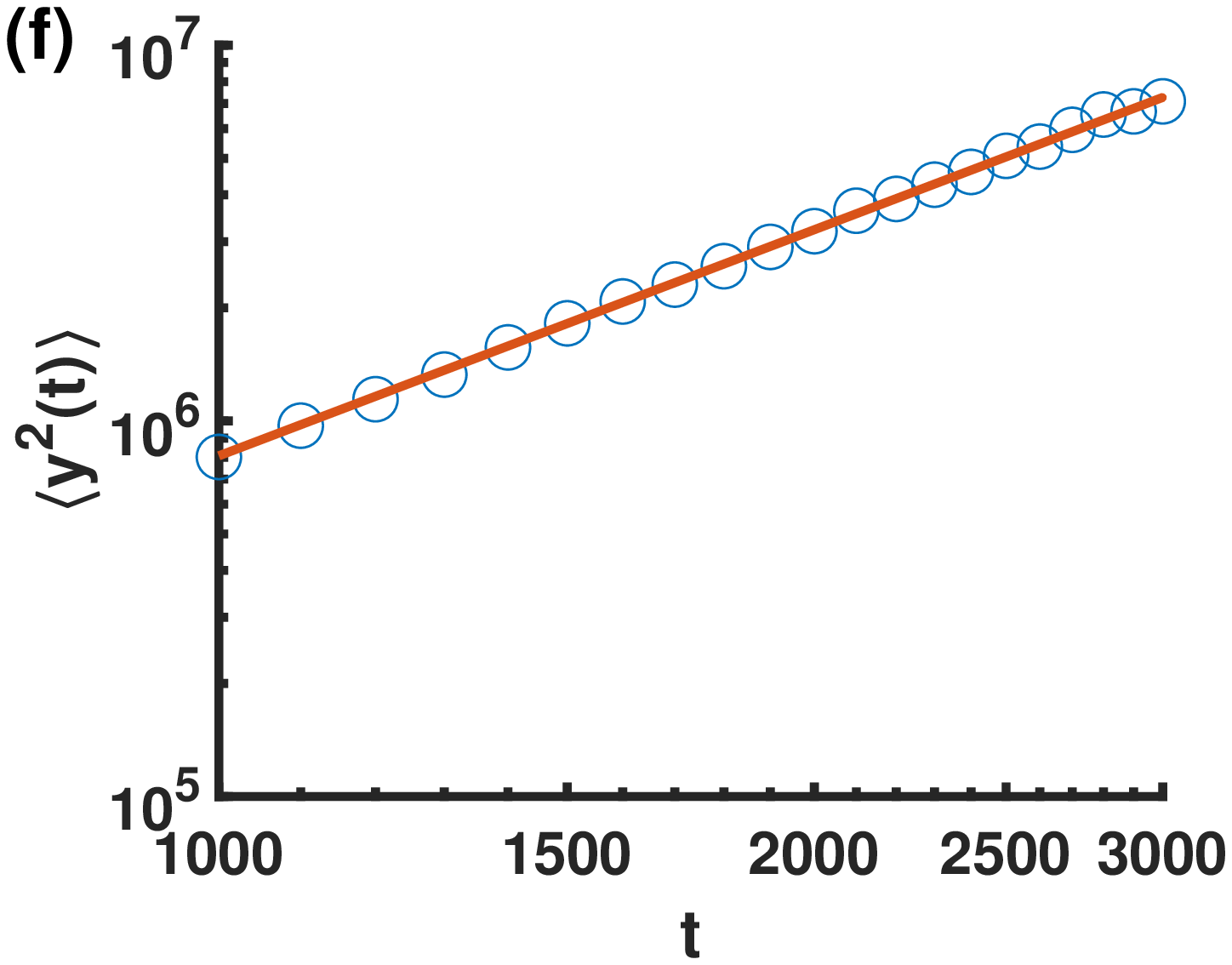}
\caption{Numerical simulations of the MSDs of L\'{e}vy walk in a uniformly power-law expansion medium by sampling over $10^4$ realizations. The walking time PDF of L\'{e}vy walk behaves as exponential distribution $\phi(\tau)=\lambda e^{-\lambda\tau}$ with $\lambda=1$. The parameter $v_0=1$. For (a) and (d), we take $t_0=1$, $\beta=0.1$ (square), and  $\beta=-0.1$ (star); for (b) and (e), we use $\beta=0.5$,    $t_0=1$ (square), and $t_0=10$ (star); for (c) and (f), we take $t_0=1$ and $\beta=1$. The solid lines in (a), (b), (c) are the theoretical results shown in \eqref{1.24}, and the solid lines in (d), (e), (f) are the theoretical results given in \eqref{1.25}.}
\label{exppowlaw}
\end{figure}

The asymptotic behaviors of the MSDs can be obtained by using  \eqref{msd}, \eqref{1.16}, \eqref{1.17}, and \eqref{1.18}, which behave as
\begin{equation}\label{1.24}
\langle x^2(t)\rangle\sim
\left\{
\begin{split}
	&\frac{2 t_0^{2\beta}v_0^2}{\lambda(1-2\beta)}t^{1-2\beta}, ~~ \mbox{if $\beta<\frac{1}{2}$, }\\
	& C_{1}+\frac{2 t_0 v_0^2 }{\lambda}\ln t, \quad \text{if}~\beta=\frac{1}{2},\\
    & C_2,\quad\quad\quad\quad\quad\quad ~ \text{if}~\beta>\frac{1}{2},
\end{split}
\right.
\end{equation}
where the constants 
$C_{1}=2 t_0 v_0^2\big(-1+e^{\lambda t_0}(\lambda t_0-1)\Gamma(0,\lambda t_0)-\ln t_0\big)/\lambda$ and $C_2=2 t_0 v_0^2 \big(2-2\beta+e^{\lambda t_0}(2\beta-1)(-2+2\beta+\lambda t_0)E_{2\beta}(\lambda t_0)\big)/[\lambda(2\beta-1)]$ with exponentially integral function 
$E_n(x)=\int_{1}^{\infty}e^{-x t}/t^n dt$ and $\Gamma(a,b)$ being the incomplete gamma function. Therefore, the asymptotic long-time behavior of the MSD in comoving coordinate displays a power-law growth with exponent $1-2\beta$ when $\beta<\frac{1}{2}$, while when $\beta=1/2$ the MSD asymptotically behaves like a logarithmic growth. Finally when $\beta>1/2$, the MSD asymptotically becomes a constant after sufficiently long time in comoving coordinate. The results of \eqref{1.24} are confirmed by the numerical simulations given in Fig. \ref{exppowlaw}(a), (b), and (c).

Correspondingly, the asymptotic behaviors of the MSDs can be obtained in physical coordinate by \eqref{1.21} for different categories of $\beta$ compared to the critical value $1/2$,
 \begin{equation}\label{1.25}
\langle y^2(t)\rangle\sim
\left\{
\begin{split}
	& \frac{2 v_0^2}{\lambda(1-2\beta)}t, ~~\quad\quad\quad \mbox{if $\beta<\frac{1}{2}$},  \\
	& C_1 t +\frac{2 t_0 v_0^2}{\lambda} t \ln t, \quad~ \mbox{if $\beta=\frac{1}{2}$}, \\
	& C_2 t^{2\beta}, \quad\quad\quad\quad\quad\quad \mbox{if $\beta>\frac{1}{2}$}.
\end{split}
\right.
\end{equation}
Intuitively, the $\beta$ controls the effects of expanding (for $\beta>0$) or contracting (for $\beta<0$) medium. As a result, for weakly expanding medium with $0<\beta<1/2$ or contracting medium, the corresponding L\'evy walk in non-static medium behaves asymptotically as a normal diffusion, which turns out to be completely different from the result of exponential scale factor shown in \eqref{1.22}, and also implies that the power-law expanding medium is much weaker than the exponential one. Besides the result of normal diffusion also indicates that the movement of the particle mainly depends on the intrinsic stochastic motion in this category of $\beta$. When the expanding effect of the medium is stronger with $\beta>\frac{1}{2}$, then the process in physical coordinate behaves like superdiffusion, which indicates expanding medium takes in charge.
%

\subsection{Power-law distributed walking time}

In the following, we consider the MSDs of L\'evy walk process in non-static medium whose walking time follows Pareto distribution
%
\begin{equation}\label{1.26}
  \phi(\tau)=\frac{1}{\tau_0}\frac{\alpha}{(1+\tau/\tau_0)^{1+\alpha}}
\end{equation}
with $\tau_0>0$ and $\alpha>0$. The corresponding asymptotic form for $\alpha\neq 1,2$ in Laplace space after neglecting higher order of small $s$ behaves like \cite{Zaburdaev}
\begin{equation}\label{waiting_time_appr}
  \hat{\phi}(s)\sim1-\frac{\tau_0}{\alpha-1}s-\tau_0^\alpha \Gamma(1-\alpha) s^\alpha+\frac{\tau_0^2 s^2}{(\alpha-1)(\alpha-2)}.
\end{equation}
Similarly, the asymptotic behaviors of the MSDs in both coordinates with respect to different types of scale factors $a(t)$ are considered in the following.

\subsubsection{Exponential scale factor}

The iteration relation of ${\widehat{T}_m(s)}$ as well as the relation between $\widehat{R}_m(s)$ and $\widehat{T}_m(s)$ for $m=0, 1, 2$ can be obtained by substituting $a(t)$ \eqref{1.19}, $\phi(\tau)$ \eqref{1.26} into \eqref{1.17}, \eqref{1.18}, respectively. Further, combining with \eqref{msd}, we have the following long time asymptotic behavior for $H>0$ in comoving coordinate
 \begin{equation}\label{1.27}
   \begin{split}
   \langle x^2(t)\rangle &\sim  \frac{\alpha v_0^2(1-\alpha-2 H \tau_0)}{4 H^2\big(1-\tau_0^\alpha2^\alpha\alpha e^{2H\tau_0}H^\alpha\Gamma(-\alpha,2 H \tau_0)\big)} \\
        & +\frac{\alpha v_0^2 e^{2 H \tau_0}(-1+\alpha)\alpha E_{1+\alpha}(2 H \tau_0)}{4 H^2\big(1-\tau_0^\alpha2^\alpha\alpha e^{2H\tau_0}H^\alpha\Gamma(-\alpha,2 H \tau_0)\big)}\\
        & +\frac{\alpha v_0^2 e^{2 H \tau_0}(4\alpha H\tau_0+4H^2\tau_0^2)E_{1+\alpha}(2 H \tau_0)}{4 H^2\big(1-\tau_0^\alpha2^\alpha\alpha e^{2H\tau_0}H^\alpha\Gamma(-\alpha,2 H \tau_0)\big)},
   \end{split}
   \end{equation}
   which is a constant.
On the other hand, for $H<0$ there exists
   \begin{equation}\label{1.28}
   \langle x^2(t)\rangle\sim
   \left\{
   \begin{split}
     \frac{1}{2}(\alpha-2)(\alpha-1)v_0^2 e^{-2 H t} t^2, & \quad \mbox{if $0<\alpha<1$},  \\
     \frac{(\alpha-1)\tau_0^{\alpha-1}v_0^2 }{(3-\alpha)} e^{-2 H t}t^{3-\alpha}, &\quad \mbox{if $1<\alpha<2$}.
   \end{split}
   \right.
   \end{equation}
For exponential expansion medium $H > 0$, the comoving variance tends to be a constant when the time is large enough, which is similar to the previous case of exponentially distributed walking time with exponential expansion. On the contrast, the MSDs display exponential growth multiplied by power-law one for exponential contraction medium $H <0$.

%

\begin{figure}[htbp]
\centering
\includegraphics[scale=0.28]{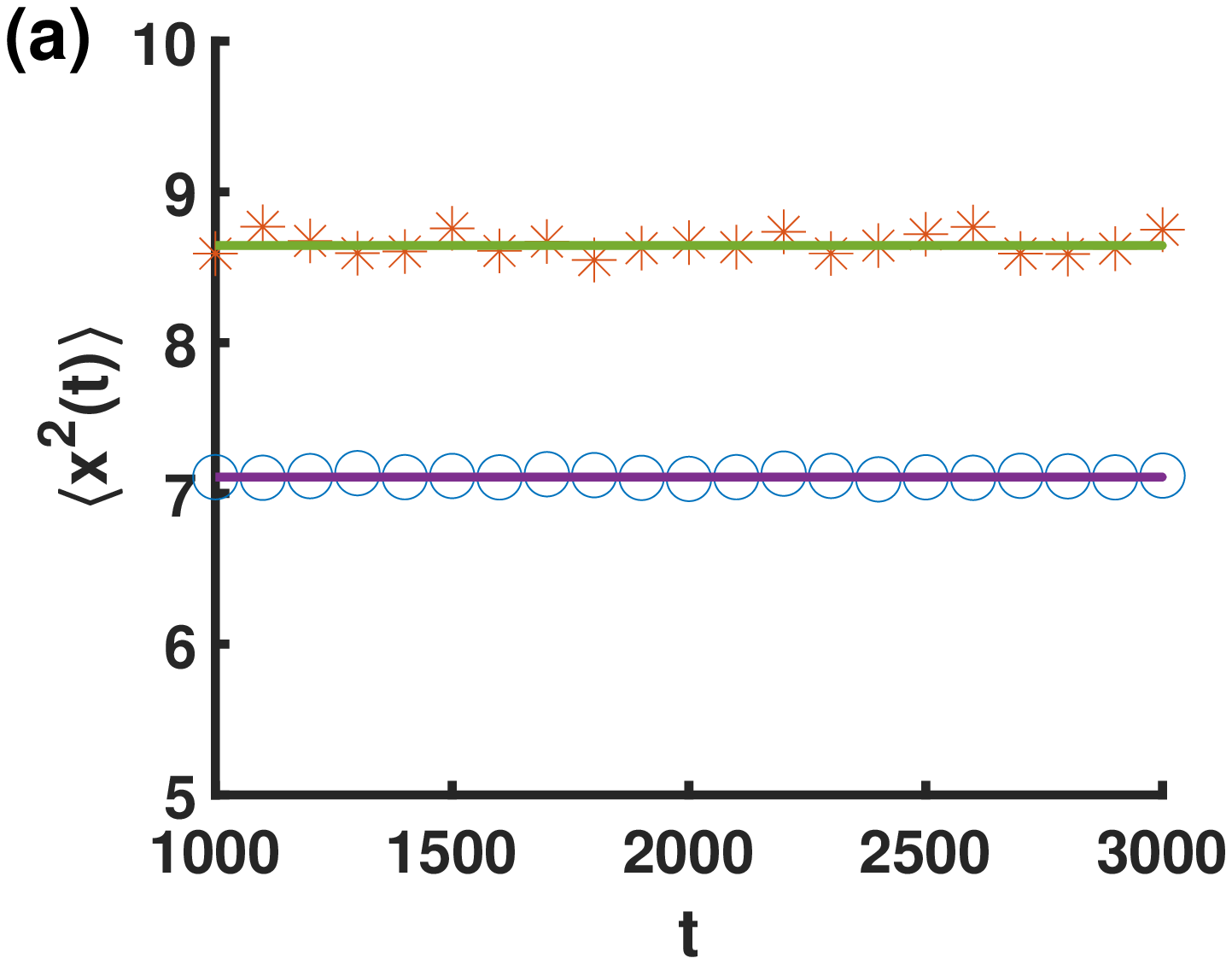}
\includegraphics[scale=0.28]{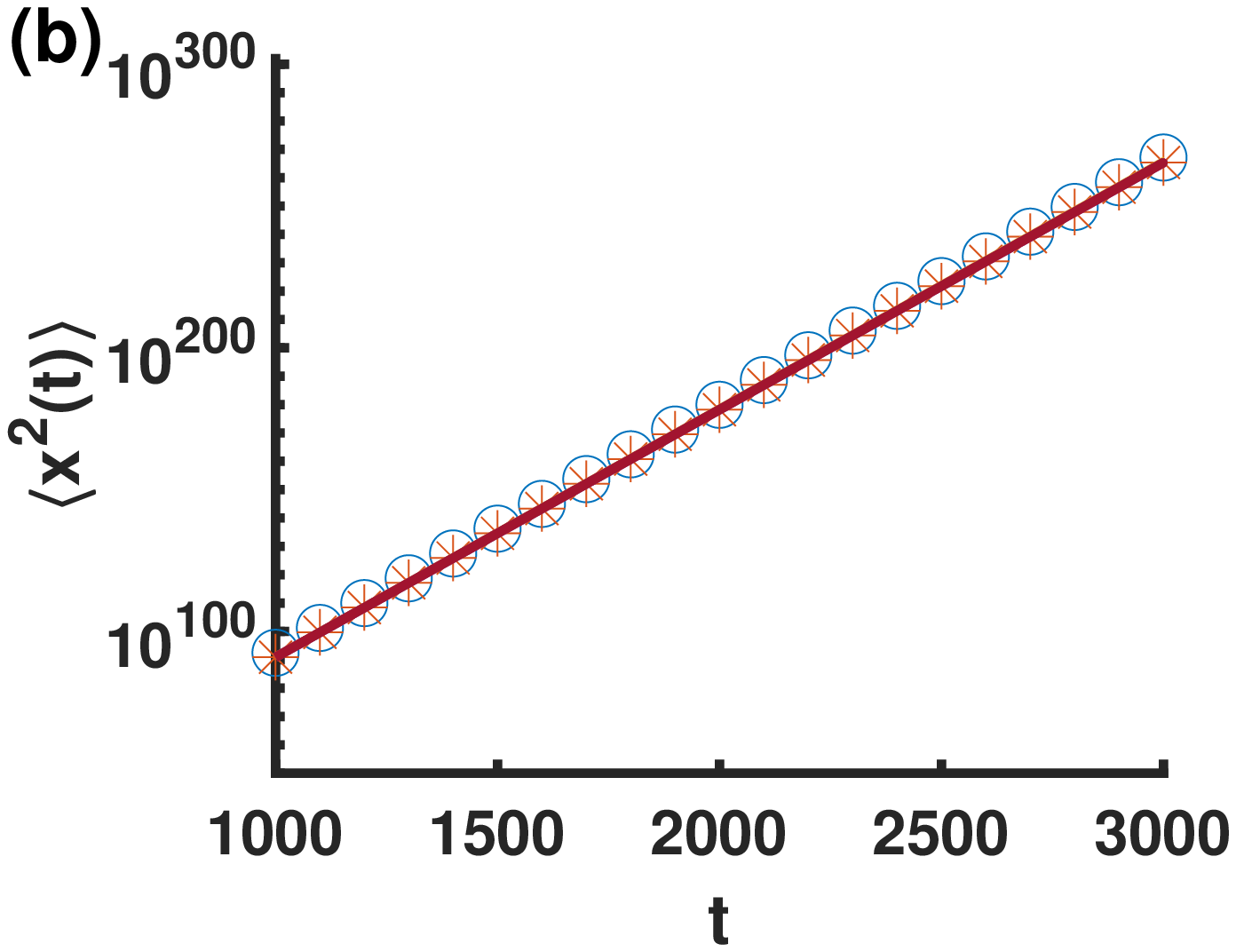}
\includegraphics[scale=0.28]{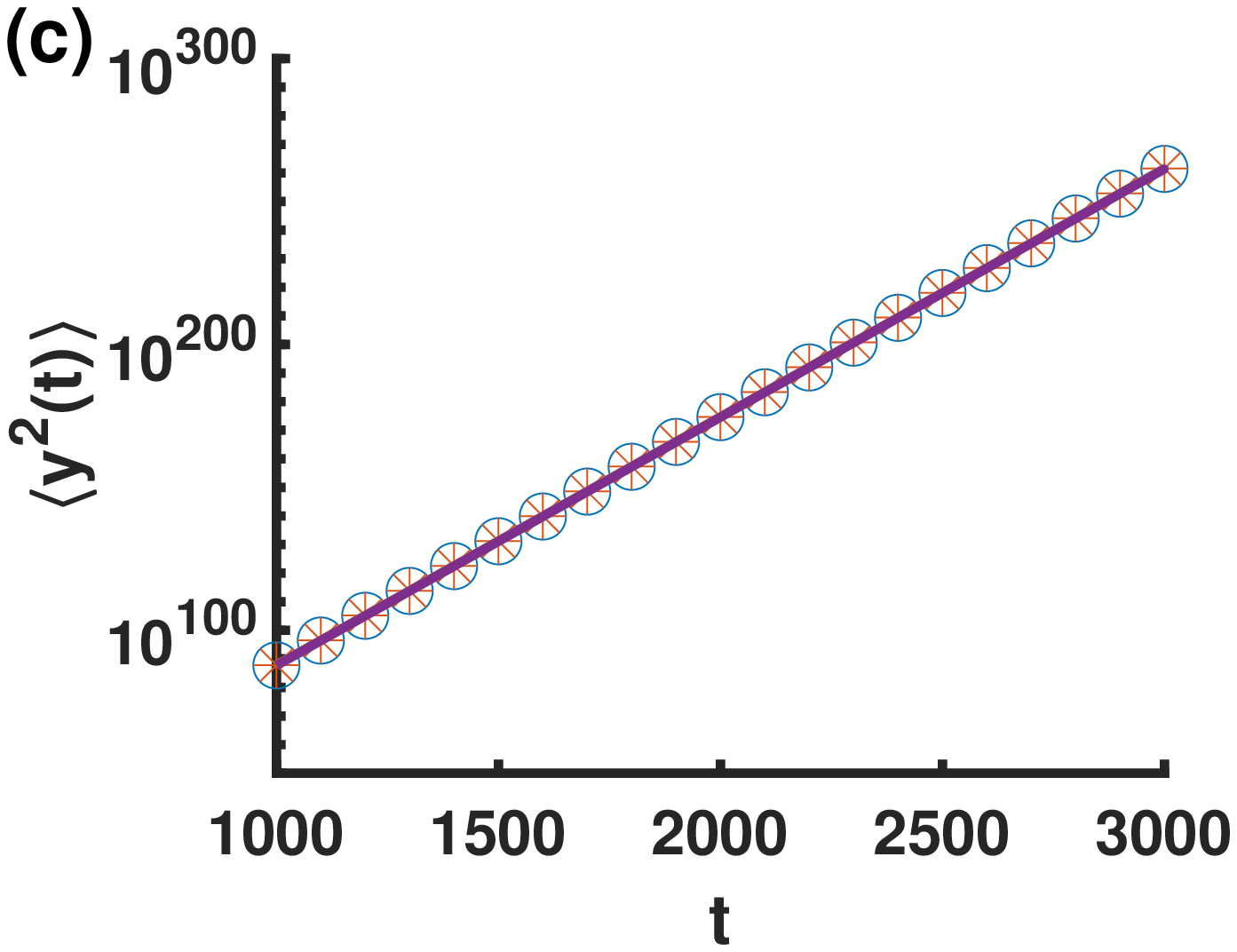}
\includegraphics[scale=0.28]{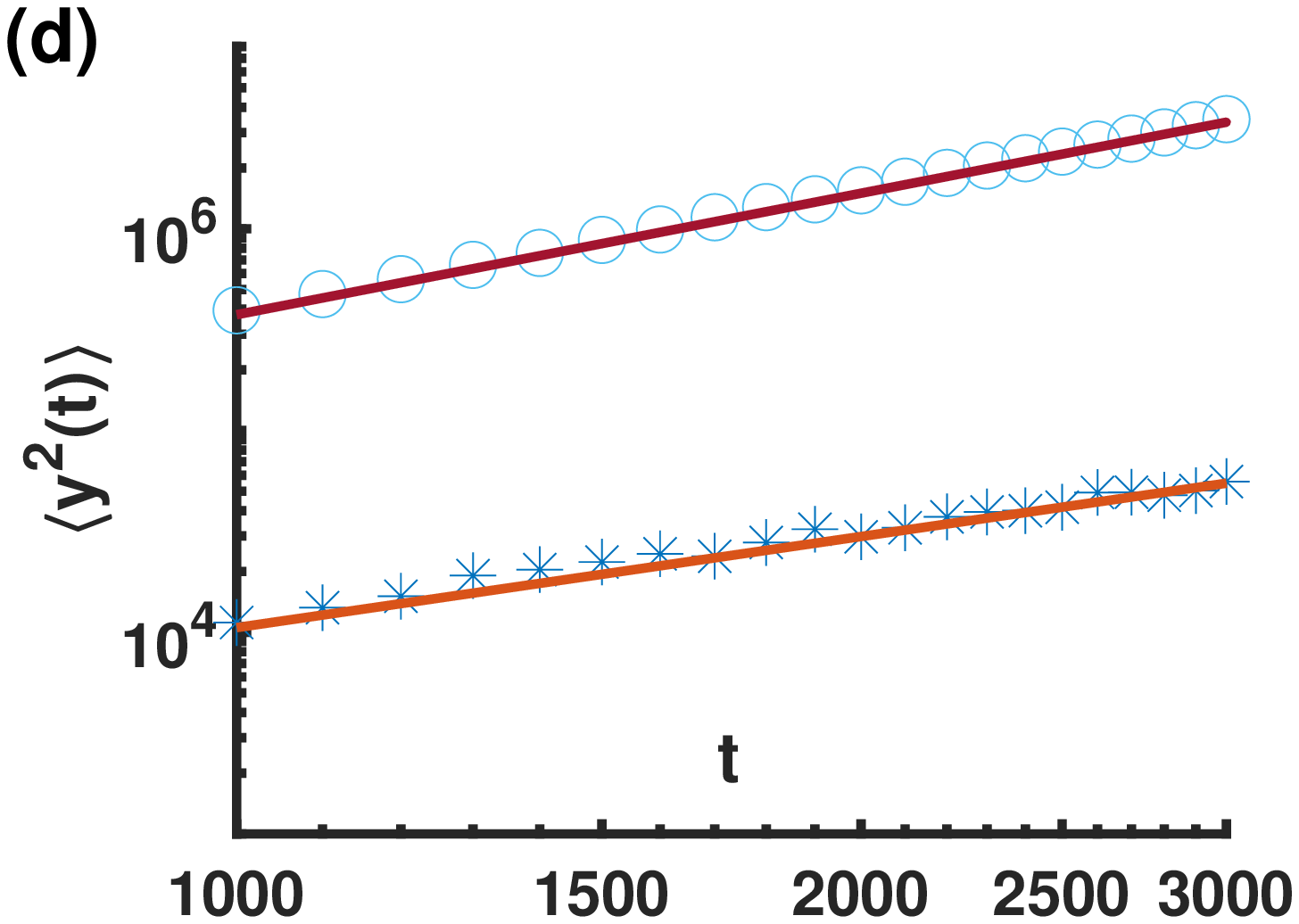}
\caption{Numerical simulations of the MSDs of L\'{e}vy walk in a uniformly exponential expansion medium by sampling over $10^4$ realizations. The walking time PDF of L\'{e}vy walk behaves as Pareto distribution with $\tau_0=1$. The parameter $v_0=1$. For (a) and (c), we take $H=10^{-1}$ and the value of $\alpha$ respectively is $\alpha=0.5$ (circles) and $\alpha=1.5$ (stars). The value of H in (b) and (d) is $H=-10^{-1}$ respectively with $\alpha=0.5$ (circles) and  $\alpha=1.5$ (stars). The solid lines in (a), (b), (c), (d) respectively are the theoretical results shown in \eqref{1.27} and \eqref{1.28} as well as \eqref{1.29} and \eqref{1.30}.}
\label{powlawexp}
\end{figure}

Furthermore, the MSDs in physical space can be figured out for sufficiently long time $t$. For $H>0$ we get
\begin{equation}\label{1.29}
\begin{split}
\langle y^2(t)\rangle&\sim \frac{\alpha v_0^2(1-\alpha-2 H \tau_0)e^{2 H t}}{4 H^2\big(1-\tau_0^\alpha2^\alpha\alpha e^{2H\tau_0}H^\alpha\Gamma(-\alpha,2 H \tau_0)\big)} \\
        & +\frac{\alpha v_0^2e^{2 H (t+\tau_0)}(-1+\alpha)\alpha E_{1+\alpha}(2 H \tau_0)}{4 H^2\big(1-\tau_0^\alpha2^\alpha\alpha e^{2H\tau_0}H^\alpha\Gamma(-\alpha,2 H \tau_0)\big)}\\
        & +\frac{\alpha v_0^2e^{2 H (t+\tau_0)}(4\alpha H\tau_0+4H^2\tau_0^2)E_{1+\alpha}(2 H \tau_0)}{4 H^2\big(1-\tau_0^\alpha2^\alpha\alpha e^{2H\tau_0}H^\alpha\Gamma(-\alpha,2 H \tau_0)\big)}.
\end{split}
\end{equation}
 And for $H<0$ we have
   \begin{equation}\label{1.30}
   \langle y^2(t)\rangle\sim \left\{
   \begin{split}
     & \frac{1}{2}(2-\alpha)(1-\alpha)v_0^2 t^2, ~\quad \mbox{if $0<\alpha<1$},  \\
     & \frac{(\alpha-1)\tau_0^{\alpha-1}v_0^2}{(3-\alpha)}t^{3-\alpha}, \quad~~ \mbox{if $1<\alpha<2$}.
   \end{split}
   \right.
   \end{equation}
A difference between the cases of positive and negative $H$ is that the MSD in physical coordinate asymptotically behaves like exponential growth for $\alpha\in (0,1)\cup (1,2)$ for $H>0$, while for the negative case the type of diffusion relies on the category of $\alpha$, specifically when $\alpha\in (0,1)$ ballistic diffusion can be observed and for $\alpha\in(1,2)$ the process behaves like superdiffusion. A significant difference for the case of $H<0$ can also be found by comparing the result of \eqref{1.30} with the localization one of \eqref{1.22}. Intuitively, a superdiffusive type of L\'evy walk in physical coordinate can keep the diffusion exponent, which is a kind of stable property mentioned in \cite{XD2018}. Further it can be found in \cite{Pengbo} that the MSD of ordinary L\'evy walk can be obtained through multiplying $\langle y^2(t)\rangle$ in \eqref{1.30} with $2/(2-\alpha)$ for $\alpha\in(0,1)\cup(1,2)$. Therefore one can conclude that the contracting exponential medium cannot localize superdiffusive or ballistic types of L\'evy walks, and it can only make the diffusion constant become slower comparing to the ordinary case instead of changing the diffusion exponent. The results of \eqref{1.29} and \eqref{1.30} are verified in Fig. \ref{powlawexp}.

\subsubsection{Power-law scale factor}

The asymptotic behaviors of the MSDs are studied when the scale factor $a(t)$ is considered to be power-law function \eqref{1.23}. Similarly, after some calculations one can obtain the asymptotic behaviors of the MSDs in comoving coordinates for different categories of $\alpha\in (0,1)\cup(1,2)$ and $\beta$. Specifically, for $0<\alpha<1$, there exists
\begin{equation}\label{1.31}
\langle x^2(t)\rangle\sim
\left\{
\begin{split}
  & \frac{(1-\alpha)\left(2+(\alpha-2)\beta\right)t_0^{2\beta}v_0^2}{2(1-\beta)} t^{2-2\beta}, \quad \mbox{if $\beta<1$},  \\
 & \alpha(1-\alpha)t_0^2v_0^2\big(\gamma+\ln t\big),\quad \mbox{if $\beta=1$}, \\
  & A_1 t^0, \quad\quad\quad\quad\quad\quad\quad\quad ~~ \mbox{if $\beta>1$};
\end{split}
\right.
\end{equation}
on the other hand for $1<\alpha<2$,
\begin{equation}\label{1.32}
\langle x^2(t)\rangle\sim
\left\{
\begin{split}
  & C_3  t^{3-\alpha-2\beta}, \quad \mbox{if $\beta<\frac{3-\alpha}{2}$ }, \\
 & \frac{\alpha(\alpha-1)t_0^{3-\alpha}\tau_0^{\alpha-1}v_0^2}{2-\alpha} \big(\gamma+\ln(t)\big), \quad\mbox{if $\beta=\frac{3-\alpha}{2}$},\\
  & A_2 t^0,\quad \mbox{if $\beta>\frac{3-\alpha}{2}$},
\end{split}
\right.
\end{equation}
where $C_3= \frac{2(\alpha-1)\left(3-\alpha+(\alpha-2)\beta\right)t_0^{2\beta}v_0^2 \tau_0^{\alpha-1}}{(6-5\alpha+\alpha^2)(3-\alpha-2\beta)}$ and $\gamma \approx 0.577216$ represents Euler–Mascheroni constant. Besides, we do not explicitly get the values of the constants $A_1$ and $A_2$ in \eqref{1.31} and \eqref{1.32} but the power-law distributed walking time  can only lead to the result of localization.

\begin{figure}[htbp]
\centering
\includegraphics[scale=0.28]{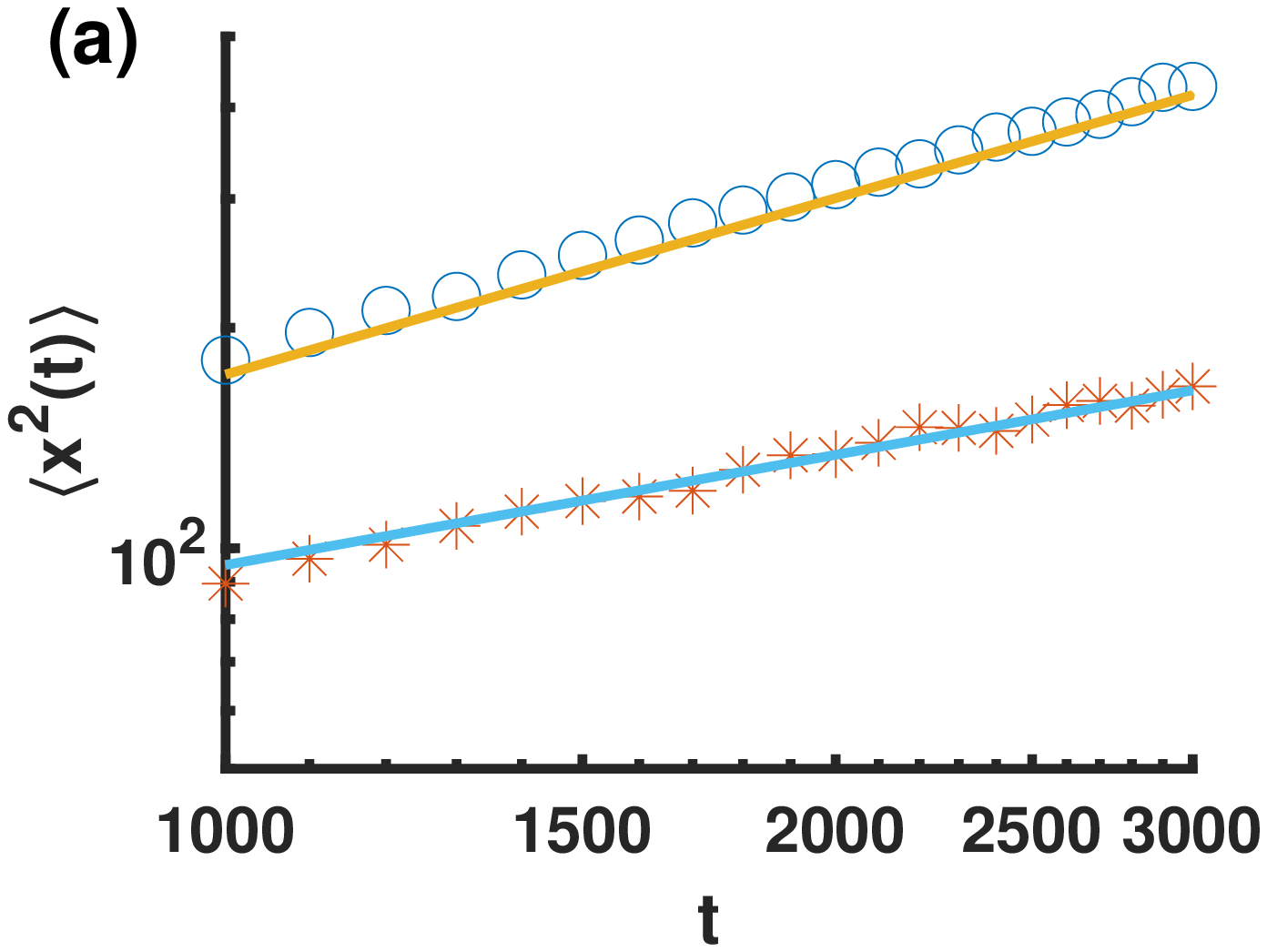}
\includegraphics[scale=0.28]{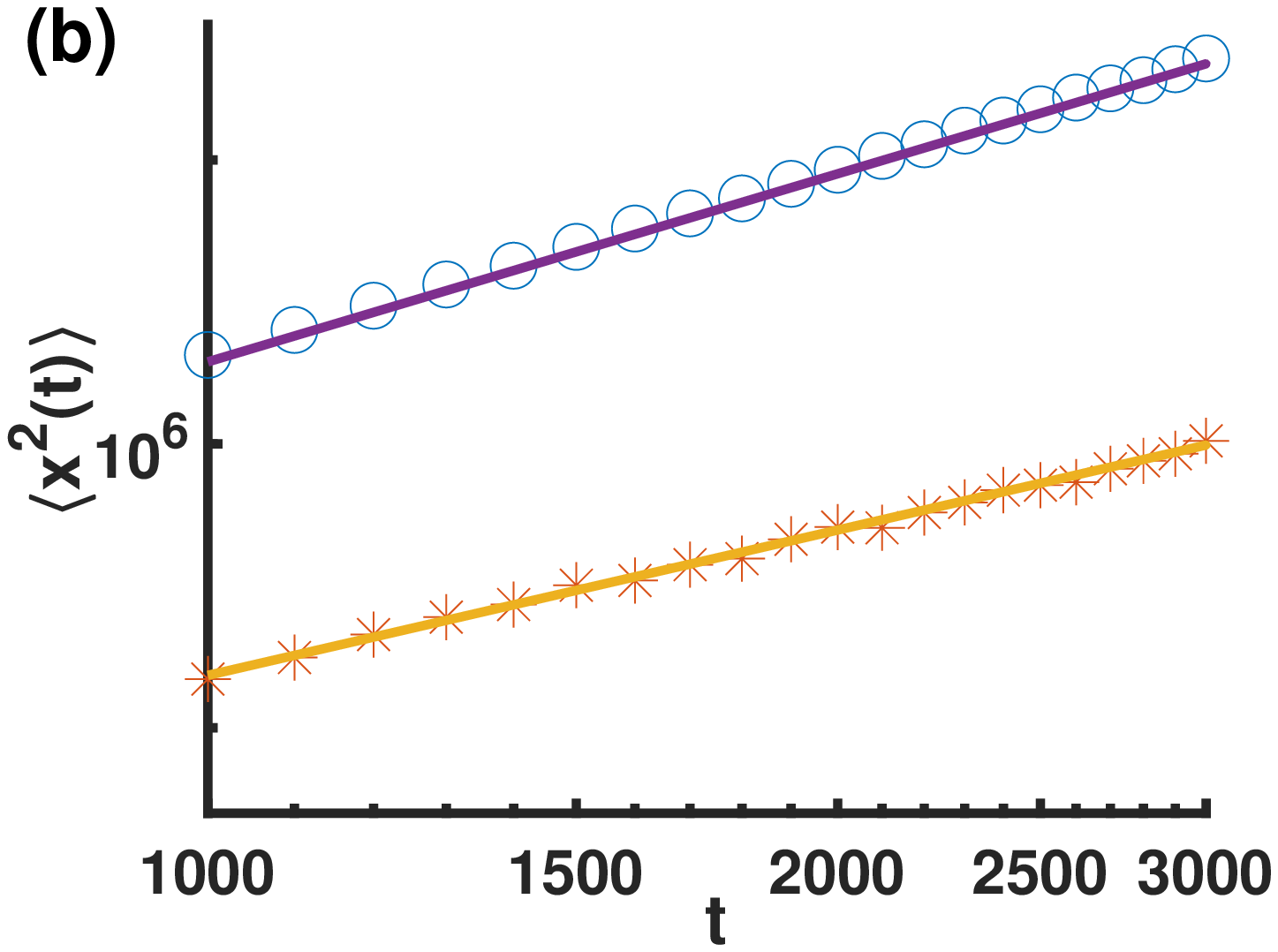}
\includegraphics[scale=0.28]{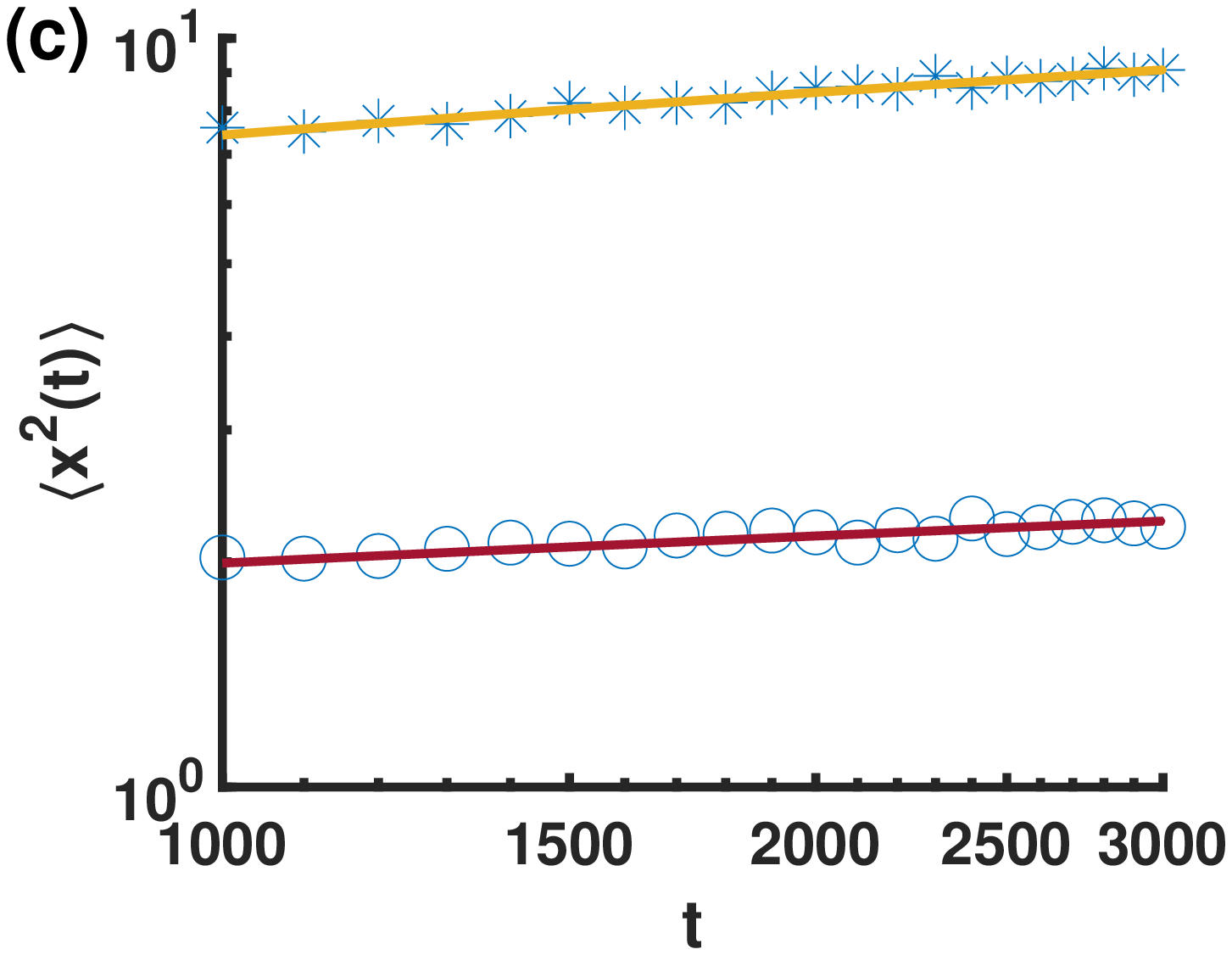}
\includegraphics[scale=0.28]{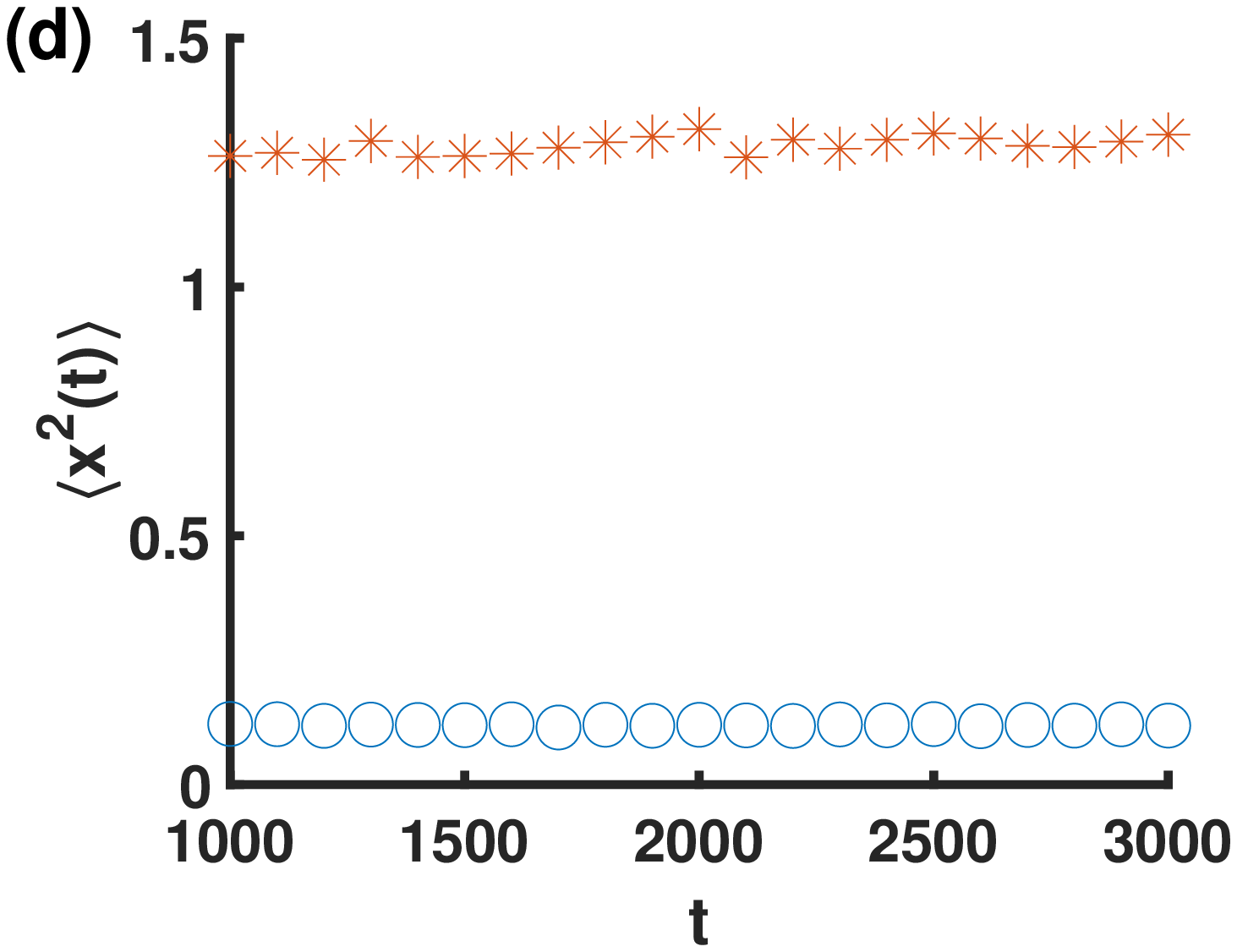}
\includegraphics[scale=0.28]{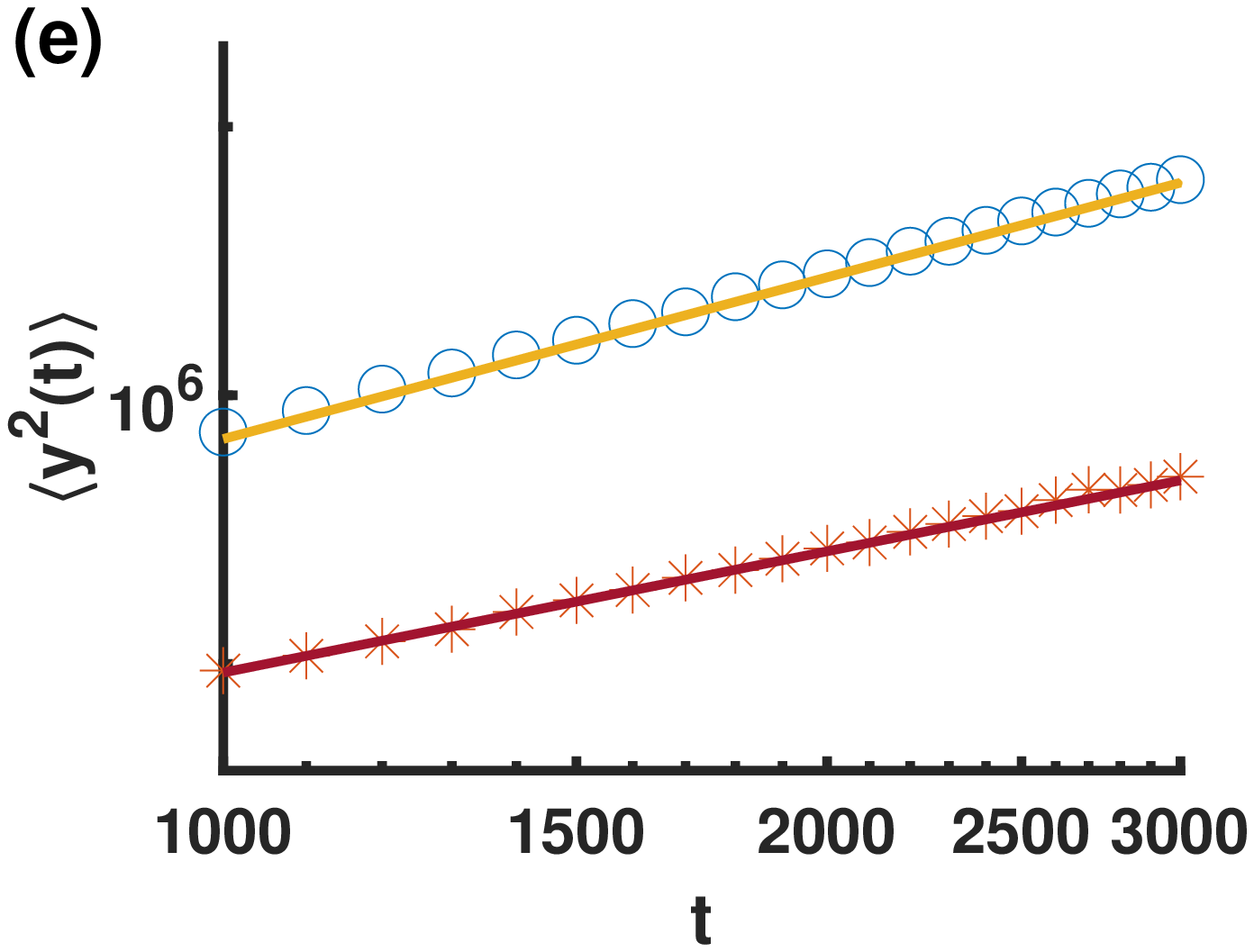}
\includegraphics[scale=0.28]{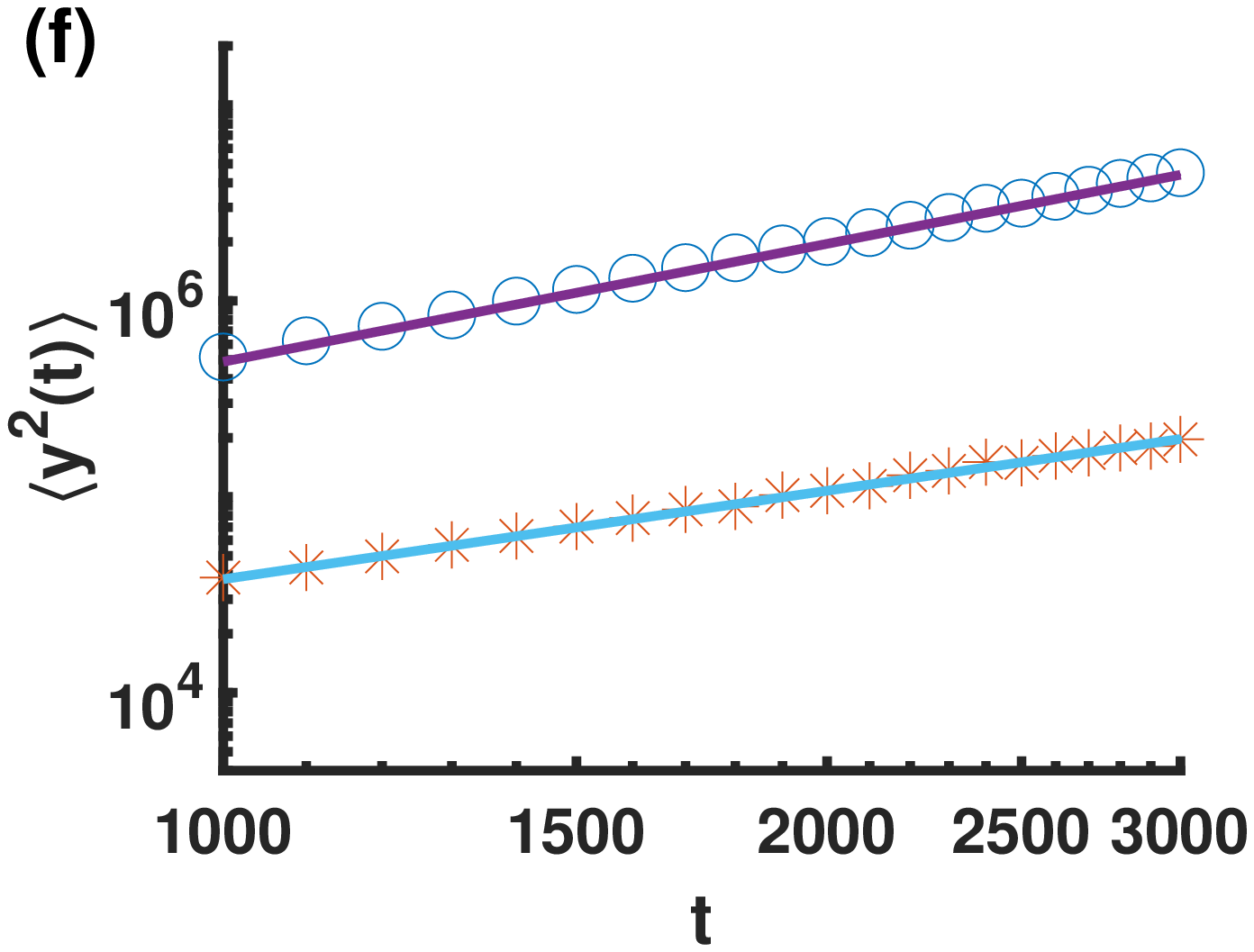}
\includegraphics[scale=0.28]{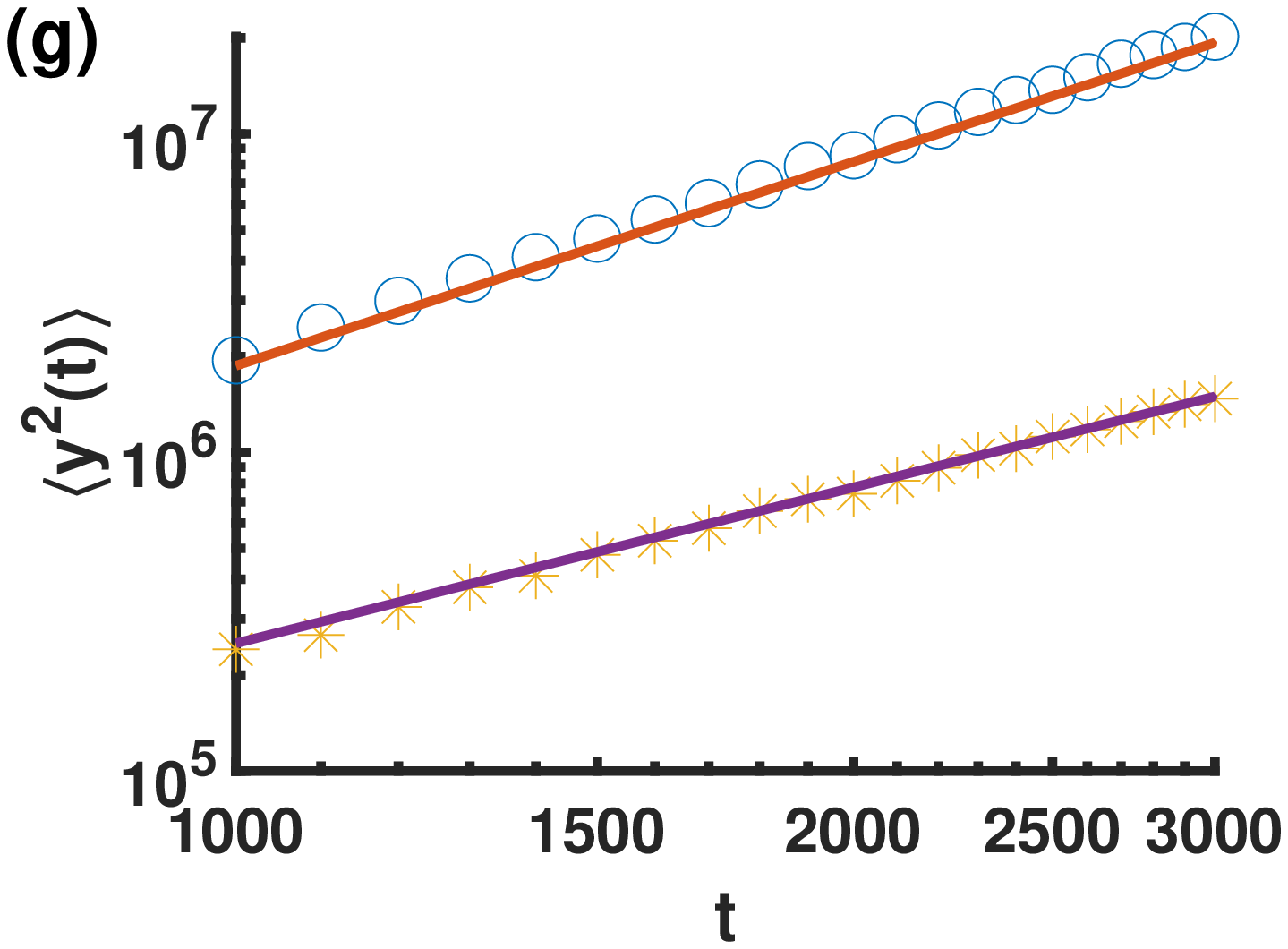}
\includegraphics[scale=0.28]{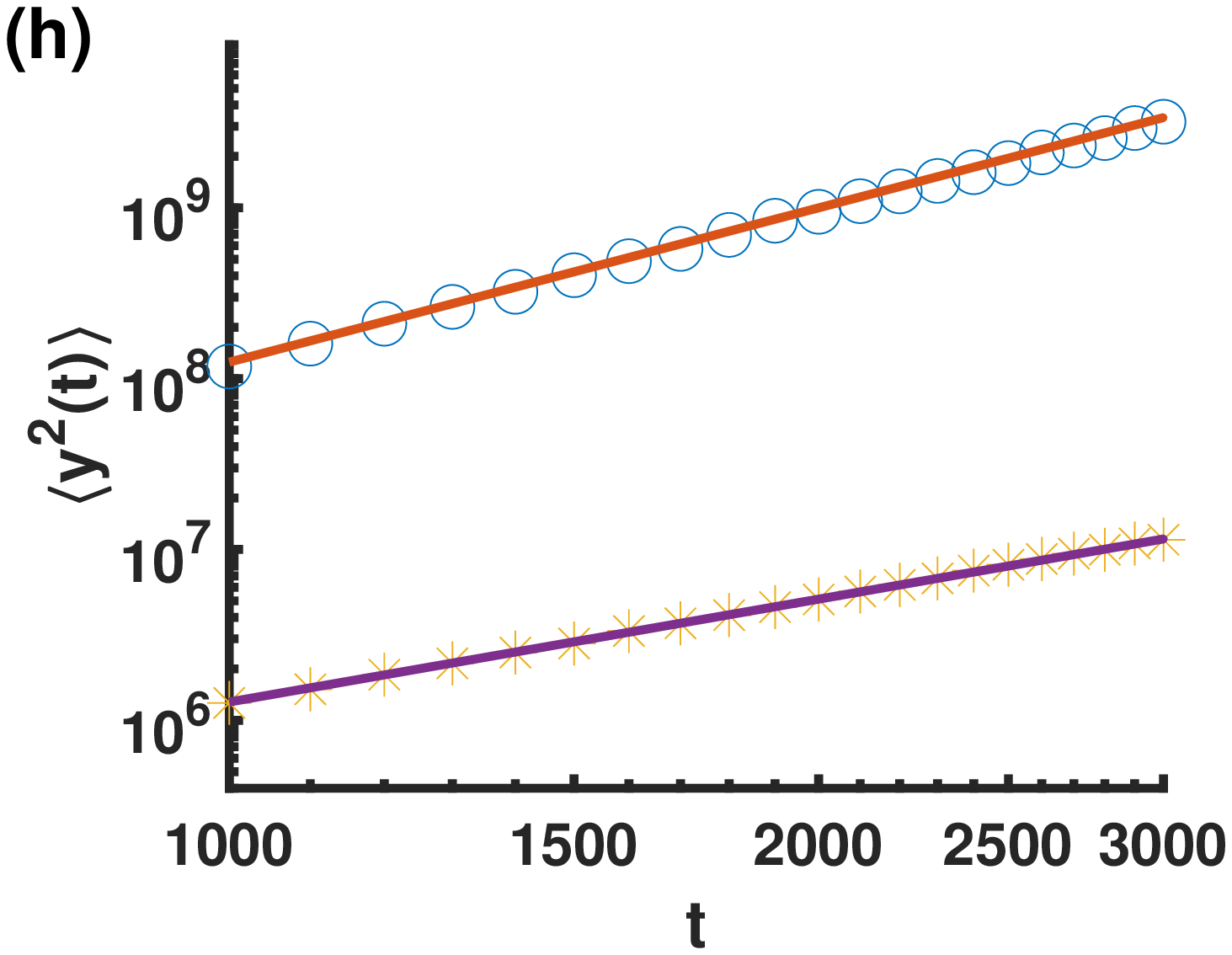}
\caption{Numerical simulations of the MSDs of L\'{e}vy walk in a uniformly power-law expansion medium by sampling over $10^4$ realizations. The walking time PDF of L\'{e}vy walk and $v_0$ are the same as the ones in Fig. \ref{powlawexp}. And $\alpha=0.5$ (circles) and $\alpha=1.5$ (stars). For $\alpha=0.5$, we respectively take $\beta=0.6$, $\beta=-0.1$, $\beta=1$, and $\beta=1.5$ in (a) (e), (b) (f), (c) (g), and (d) (h); for $\alpha=1.5$, we respectively use $\beta=0.5$, $\beta=-0.1$, $\beta=0.75$, and $\beta=1$ in (a) (e), (b) (f), (c) (g), and (d) (h). The solid lines in (a), (b) and (c), (d) are the theoretical results shown in \eqref{1.31} and \eqref{1.32} and the solid lines in (e), (f) and (g), (h) are the theoretical results shown in \eqref{1.311} and \eqref{1.322}.}
\label{powlawpowlaw}
\end{figure}

Similarly, from the relation between comoving and physical coordinates \eqref{1.21}, the MSDs in physical coordinates can be obtained from \eqref{1.31} and \eqref{1.32}. Specifically, for $0<\alpha<1$,
\begin{equation}\label{1.311}
\langle y^2(t)\rangle\sim
\left\{
\begin{split}
  & \frac{(1-\alpha)\left(2+(\alpha-2)\beta\right)v_0^2}{2(1-\beta)} t^{2}, \quad \mbox{if $\beta<1$ },  \\
  & \alpha(1-\alpha)v_0^2\big(\gamma+\ln(t)\big)t^{2}, \quad \mbox{if $\beta=1$},\\
  & A_1\left(\frac{t}{t_0}\right)^{2\beta}, \quad\quad\quad\quad\quad\quad~ \mbox{if $\beta>1$};
\end{split}
\right.
\end{equation}
and for $1<\alpha<2$,
\begin{equation}\label{1.322}
\langle y^2(t)\rangle\sim
\left\{
\begin{split}
  & C_3 t^{3-\alpha}, \quad \mbox{if $\beta<\frac{3-\alpha}{2}$ },  \\
  & \frac{\alpha(\alpha-1)\tau_0^{\alpha-1}v_0^2\big(\gamma+\ln(t)\big)}{2-\alpha}t^{3-\alpha},\quad \mbox{if $\beta=\frac{3-\alpha}{2}$},\\
  & A_2\left(\frac{t}{t_0}\right)^{2\beta}, \quad \mbox{if $\beta>\frac{3-\alpha}{2}$}.
\end{split}
\right.
\end{equation}
The MSDs obtained in both coordinates are verified in Fig. \ref{powlawpowlaw}. As we can observe from \eqref{1.311} and \eqref{1.322}, there exists a critical value of $\beta$ denoted as $\beta_c$ such that when $\beta<\beta_c$ the L\'evy walk in physical coordinate can keep the ballistic diffusion or superdiffusion, whereas for $\beta>\beta_c$ the MSDs in physical coordinates asymptotically behave like $\langle y^2(t)\rangle\sim t^{2\beta}$ for $\alpha\in(0,1) \cup (1,2)$. Therefore, when $\beta>\beta_c$ the expanding medium will play a key role to accelerate the L\'evy walk process in physical coordinate, and when $\beta<\beta_c$  the effect of non-static medium will become weaker so that the L\'evy walk  can keep the diffusion exponent. The specific value of $\beta_c$ relies on the categories of $\alpha$; if $\alpha\in (0,1)$ then $\beta_c=1$ and if $\alpha\in (1,2)$ then $\beta_c=(3-\alpha)/2$. When $\beta=\beta_c$, a logarithmic correction of $t$ is needed.

%

\section{Mean First passage time}\label{sec 5}

In this section, we mainly consider the asymptotic behaviors of mean first passage time $\langle t_f\rangle$, which has wide applications in many different areas. The first passage time $t_f$ is a random time when a stochastic process reaches the spatial boundary for the first time. Since the  non-static medium causes the physical space always to expand or contract, the spatial boundary also changes correspondingly. Therefore it is reasonable to consider L\'evy walk process moving in a given domain of comoving space. Specifically, we consider the one dimensional L\'evy walk process moving in an interval $[-L,L]$ of comoving space with $L>0$ and absorbing boundary $\pm L$. Additionally, we choose the scale factor $a(t)$ to be power-law function \eqref{1.23}. According to the numerical simulations with respect to different types of PDFs $\phi(\tau)$ of walking time, we present the asymptotic behaviors of $\langle t_f\rangle$ as a function of $L$
%

\begin{figure}[htbp]
\centering
\includegraphics[width=8cm]{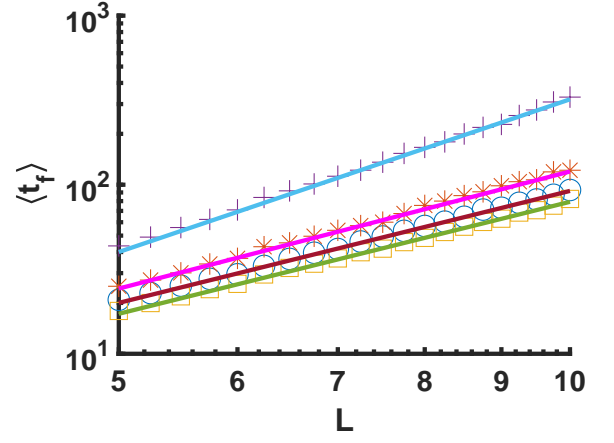}
\caption{Numerical simulations of MFPT of L\'{e}vy walk in comoving coordinate where the physical space is doing power-law expansion. The results are obtained by sampling over $10^4$ realizations. The walking time PDF of L\'{e}vy walk behaves as exponential distribution $\phi(\tau)=\lambda e^{-\lambda\tau}$. The parameters are $v_0=1$ and $x_0=0$. For circles, the parameters are $\lambda=1$, $\beta=0.1$, and $t_0=1$; for stars,  $\lambda=2$, $\beta=0.1$, and $t_0=1$; for squares, $\lambda=1$, $\beta=0.1$, and $t_0=2$; for plus, $\lambda=1$, $\beta=0.2$, and $t_0=1$. The solid lines represent the result of $\langle t_f\rangle \sim L^{-\frac{\lambda}{10}+8\beta+\frac{13}{10}}$.}
\label{mfptexp}
\end{figure}

\begin{figure}[htbp]
\centering
\includegraphics[scale=0.28]{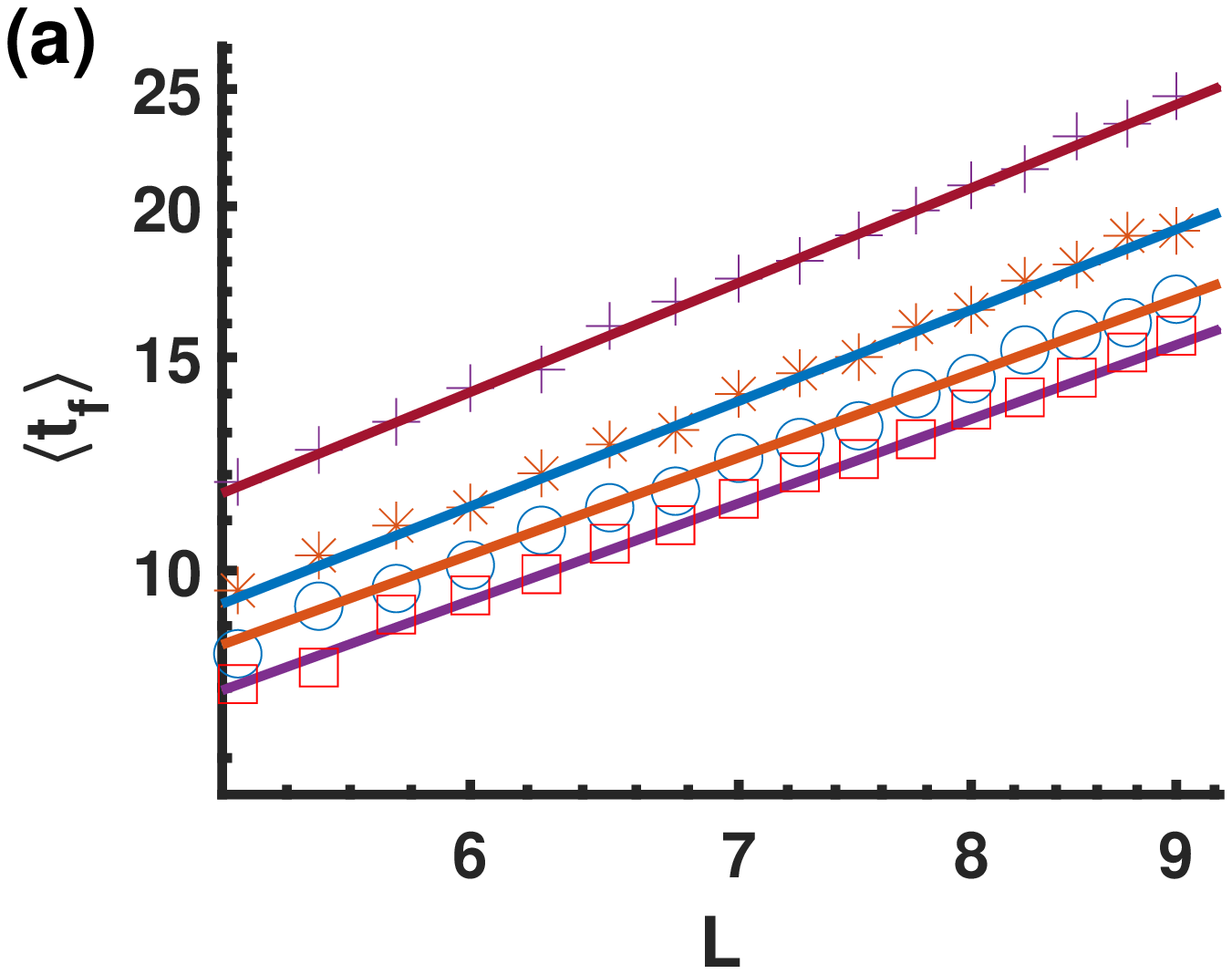}
\includegraphics[scale=0.28]{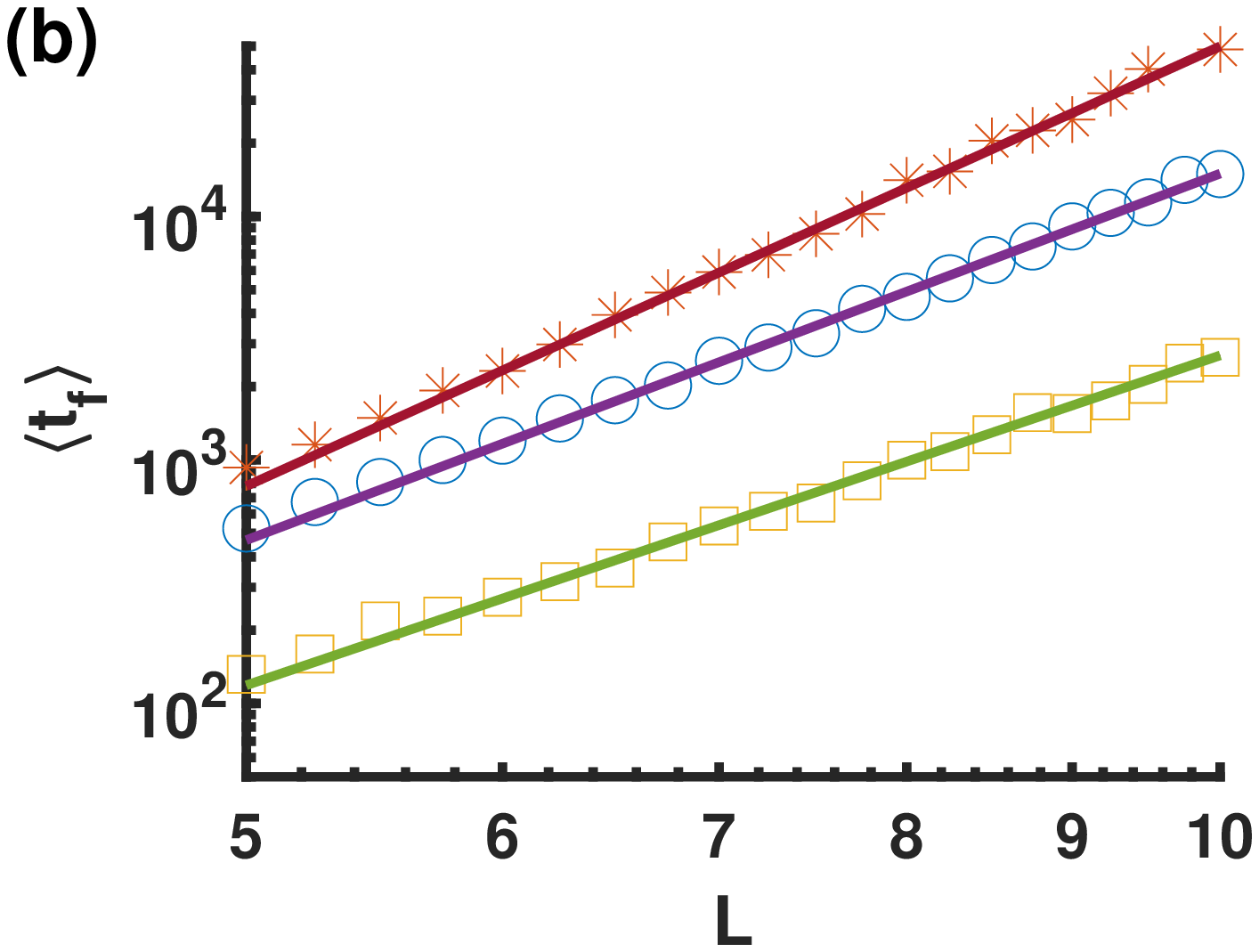}
\caption{Numerical simulations of MFPT of L\'{e}vy walk in comoving coordinate where the physical space is doing the  power-law expansion. The results are obtained by sampling over $10^4$ realizations. The walking time PDF of L\'{e}vy walk behaves as power-law distribution $\phi(\tau)=\frac{1}{\tau_0}\frac{\alpha}{(1+\tau/\tau_0)^{1+\alpha}}$. The parameters are $v_0=1$ and $x_0=0$. The panel (a) is with $0<\alpha<1$.  For circles, the parameters are  $\alpha=0.5$, $\beta=0.1$, and $t_0=1$; for star,  $\alpha=0.7$, $\beta=0.1$, and $t_0=1$; for squares, $\alpha=0.5$, $\beta=0.1$, and $t_0=2$; for plus, $\alpha=0.5$, $\beta=0.2$, and $t_0=1$. The panel (b) is with $1<\alpha<2$. For circles, the parameters are  $\alpha=1.5$, $\beta=0.5$, and $t_0=1$; for stars,  $\alpha=1.6$, $\beta=0.5$, and $t_0=1$; for squares,  $\alpha=1.5$, $\beta=0.5$, and $t_0=2$. The solid lines respectively represent the results of $\langle t_f\rangle \sim L^{\frac{\alpha}{2}+\frac{3}{2}\beta+\frac{4}{5}}$ and $\langle t_f\rangle \sim L^{10 \alpha+15 \beta-\frac{t_0}{2}-17}$.}
\label{mfptpower}
\end{figure}

When the walking time PDF of L\'{e}vy walk is exponential distribution, i.e., $\phi(\tau)=\lambda e^{-\lambda\tau}$, the asymptotic behavior of the mean first passage time for large $L$ is given by numerical simulation in Fig. \ref{mfptexp}, which matches
\begin{equation}\label{mfpt1}
  \langle t_f\rangle\sim L^{-\frac{\lambda}{10}+8\beta+\frac{13}{10}}.
\end{equation}
For L\'{e}vy walk with power-law walking time $\phi(\tau)=\frac{1}{\tau_0}\frac{\alpha}{(1+\tau/\tau_0)^{1+\alpha}}$, when $0<\alpha<1$, through numerical simulation we find
\begin{equation}\label{mfpt2}
  \langle t_f\rangle\sim L^{\frac{\alpha}{2}+\frac{3}{2}\beta+\frac{4}{5}}.
\end{equation}
Similarly, when $1<\alpha<2$, by numerical simulation we have
\begin{equation}\label{mfpt3}
  \langle t_f\rangle\sim L^{10 \alpha+15 \beta-\frac{t_0}{2}-17}.
\end{equation}
The above results are verified by Fig. \ref{mfptpower}.

\section{Conclusion}\label{sec 6}

In this paper, we establish the model of L\'evy walk in non-static medium and study the stochastic dynamics based on the model. 
We first build the governing equation for the PDF of the position of the particle in comoving coordinate. Further, we utilize Hermite orthogonal polynomials to approach the solution to governing equation, which can also lead to the form of the MSD in comoving coordinate. According to the time dependent scale factor, we build up the relation of the MSD in comoving coordinate with the one in physical coordinate. Next we give some representative examples of scale factors $a(t)$ and walking time density $\phi(\tau)$ to obtain the statistical properties in both coordinates through analytical analyses and numerical simulations.
%

For L\'evy walk with exponentially distributed walking time, we first consider the case of the scale factor $a(t)$ to be an exponential function. When the Hubble constant $H>0$ indicating an expanding medium, the MSD in comoving coordinate tends to be a constant, leading to exponential growth in physical coordinate. On the contrary for the case of contracting medium with $H<0$, we reach a different conclusion, specifically, the MSDs in comoving and physical coordinates show exponential growth and a constant, respectively. The results also indicate that the non-static medium with exponential scale factor has significant influence on the L\'evy walk whose walking time obeys exponential PDF with average $1/\lambda$. Additionally we simulate the stationary distributions for the processes with $H>0$ in comoving coordinate and $H<0$ in the physical coordinate, it turns out that a transition from unimodal distribution to the bimodal one can be observed with the decreasing of $\lambda$ for the former case or increasing of absolute value of the Hubble constant $|H|$ for the latter case. Next, the kurtosis $K$ which is also an important quantity for stochastic process has been theoretically studied in physical coordinate for $H<0$. According to the simulation, we find $K$ is a function of $H$ and $\lambda$ for sufficiently long time, and the evolutions of $K$ for each variable are analyzed. Then we consider the case that the scale factor is a power-law function; by calculating and analyzing the MSD $\langle y^2(t)\rangle$ in physical coordinate we find that for $\beta<1/2$ there is $\langle y^2(t)\rangle\sim t$ indicating the domination of L\'evy walk process comparing to non-static medium, whereas when $\beta>1/2$ it is $\langle y^2(t)\rangle \sim t^{2\beta}$ indicating that the expending medium dominates the process.


Next for L\'{e}vy walk with a Pareto distributed walking time with $\alpha\in(0,1)\cup (1,2)$, we also consider exponential scale factor first. When $H>0$ so that the medium expands, the constant MSD in comoving coordinate implies exponential growth of the MSD in physical coordinate; and for contracting medium with $H<0$, the MSD in physical coordinate keeps the same diffusion exponential as the ordinary L\'evy walk in static medium, and the contracting medium in this case can only shrink the value of diffusion constant. Therefore for the exponential scale factor, one can conclude that the expanding medium with $H>0$ can significantly affect L\'evy walk process whereas the contracting medium has negligible influence on the diffusion. Finally for the case of power-law scale factor, a critical value $\beta_c$ is found; when $ \beta<\beta_c$ the MSD in physical coordinate keeps the same diffusion exponent as the ordinary case in static medium. For different region of $\alpha$, the critical value is also different, specifically, $\beta_c=1$ and $(3-\alpha)/2$ for $0<\alpha<1$ and $1<\alpha<2$, respectively. Besides, the asymptotic behaviors of mean first passage time in comoving coordinate are also discussed through numerical simulations.
%

\section*{Acknowledgements}

This work was supported by the National Natural Science Foundation of China under Grant No. 12071195, and the AI and Big Data Funds under Grant No. 2019620005000775.

\begin{appendix}

\section{A brief introduction of Hermite polynomials}\label{Appen_A}

Hermite polynomials are a set of orthogonal polynomials defined on $(-\infty,\infty)$ with weight function $e^{-x^2}$ \cite{hermit_intro}. One way of standardizing the Hermite polynomials is to explicitly give them as
\begin{equation}\label{a1}
  H_n(x)=(-1)^n e^{x^2} \frac{d^n}{d x^n} e^{-x^2}.
\end{equation}
Furthermore, its orthogonality can be expressed as
\begin{equation}\label{a2}
  \int_{-\infty}^{\infty} H_n(x)H_m(x)e^{-x^2}dx=\sqrt{\pi}2^n n! \delta_{n,m},
\end{equation}
where $\delta_{n,m}$ is the Kronecker delta function.
By Taylor's expansion, there exists
\begin{equation}\label{a3}
  H_n(x+y)=\sum_{k=0}^{n}\binom{n}{k} H_k(y) (2 x)^{n-k};
\end{equation}
and the following holds
\begin{equation}\label{a4}
  H_n(\gamma x)=\sum_{j=0}^{\lfloor\frac{n}{2}\rfloor} \gamma^{n-2 j} (\gamma^2-1)^j \binom{n}{2 j} \frac{(2 j)!}{j!}H_{n-2 j}(x),
\end{equation}
where $\lfloor\frac{n}{2}\rfloor$ is the biggest integer smaller than $\frac{n}{2}$. The Hermite polynomials  evaluated at zero, 
called Hermite number, are
\begin{equation}\label{a5}
  H_n(0)=
  \begin{cases}
       0, &\mbox{if $n$ is odd}, \\
       (-1)^{\frac{n}{2}} 2^{\frac{n}{2}} (n-1)!!, & \mbox{if $n$ is even}.
     \end{cases}
\end{equation}

In particular,
\begin{equation}\label{a6}
   H_0(x)=1,\quad H_1(x)=2 x,\quad {\rm and}~~  H_2(x)=4 x^2-2.
\end{equation}

\section{Derivations of \eqref{1.11} and \eqref{1.12}}\label{App_B}

Inserting \eqref{1.9} into \eqref{1..6} leads to
\begin{widetext}
\begin{equation}\label{B1}
  \begin{split}
     \sum_{n=0}^{\infty} H_n(x)T_n(t) e^{-x^2}&=\frac{1}{2}\int_{0}^{t}\sum_{n=0}^{\infty} H_n\bigg(x-\frac{v_0 \tau}{a(t)} \bigg)
      T_n(t-\tau) \exp\Bigg[-\bigg(x-\frac{v_0 \tau}{a(t)}\bigg)^2\Bigg] \phi(\tau) d\tau\\
      & +\frac{1}{2}\int_{0}^{t}\sum_{n=0}^{\infty} H_n\left(x+\frac{v_0 \tau}{a(t)}\right)
      T_n(t-\tau) \exp\left[-\left(x+\frac{v_0 \tau}{a(t)}\right)^2\right] \phi(\tau) d\tau
       +P_0(x)\delta(t).
  \end{split}
\end{equation}
Multiplying $H_m(x)$, $m=0,1,2,\cdots,$ on both side of \eqref{B1} and integrating $x$ over $(-\infty,+\infty)$,  according to the orthogonal properties of Hermite polynomials \eqref{a2}, the left hand of \eqref{B1} becomes
\begin{equation}\label{B2}
 \int_{-\infty}^{+\infty}\sum_{n=0}^{\infty} H_n(x)T_n(t)H_m(x) e^{-x^2} dx=\sqrt{\pi}2^m m! T_m(t).
\end{equation}
We next pay attention to the first term of the right hands of \eqref{B1} in detail. Multiplying $H_m(x)$, $m=0,1,2,\cdots,$ and then integrating $x$ over $(-\infty,+\infty)$ lead to
\begin{equation}\label{B3}
\begin{split}
      &\frac{1}{2}\int_{0}^{t} \phi(\tau) T_n(t-\tau) \int_{-\infty}^{+\infty} H_m(x)\sum_{n=0}^{\infty} H_n\left(x-\frac{v_0 \tau}{a(t)}\right)
     \exp\left[-\left(x-\frac{v_0 \tau}{a(t)}\right)^2\right] dx d\tau \\
      =& \frac{1}{2}\int_{0}^{t} \phi(\tau) T_n(t-\tau) \int_{-\infty}^{+\infty} H_m\left(y+\frac{v_0 \tau}{a(t)}\right)\sum_{n=0}^{\infty} H_n(y)
     \exp(-y^2) dy d\tau \\
     =&\frac{1}{2}\int_{0}^{t} \phi(\tau) T_n(t-\tau) \int_{-\infty}^{+\infty}\sum_{k=0}^{m}\binom{m}{k} H_k(y) \left(\frac{2v_0 \tau}{a(t)}\right)^{m-k}\sum_{n=0}^{\infty} H_n(y)
     \exp(-y^2) dy d\tau \\
     =&\frac{1}{2}\sum_{k=0}^{m}\binom{m}{k} \int_{0}^{t} \phi(\tau) T_n(t-\tau)\left(\frac{2v_0 \tau}{a(t)}\right)^{m-k} \int_{-\infty}^{+\infty} H_k(y)\sum_{n=0}^{\infty} H_n(y)
     \exp(-y^2) dy d\tau \\
     =&\frac{1}{2}\sum_{k=0}^{m}\frac{m!}{k!(m-k)!} \int_{0}^{t} \sqrt{\pi}2^k k! \left(\frac{2v_0 \tau}{a(t)}\right)^{m-k}T_k(t-\tau) \phi(\tau) d\tau,
\end{split}
\end{equation}
which is obtained by utilizing the properties of Hermite polynomials \eqref{a2} and \eqref{a3} in Appendix \ref{Appen_A}. The second term can be treated in the similar way and it reduces to  $\frac{1}{2}\sum_{k=0}^{m}\frac{m!}{k!(m-k)!} \int_{0}^{t} \sqrt{\pi}2^k k! \left(\frac{-2v_0 \tau}{a(t)}\right)^{m-k}T_k(t-\tau) \phi(\tau) d\tau$. Since the initial distribution of particles is Dirac-delta function, i.e., $P_0(x)=\delta(x)$, the third term becomes $\int_{-\infty}^{+\infty}H_m(x)\delta(x)\delta(t)dx=\delta(t)H_m(0)$.

Finally the iteration relation of ${T_m(t)}$ can be obtained 
\begin{equation}\label{B4}
 \sqrt{\pi} 2^m m!T_m(t)= \frac{1}{2}\sum_{k=0}^{m}\frac{m!}{k!(m-k)!}\int_{0}^{t}\sqrt{\pi}2^k k!\left[\left(\frac{2 v_0 \tau}{a(t)}\right)^{m-k}+\left(\frac{-2 v_0 \tau}{a(t)}\right)^{m-k}\right]T_k(t-\tau)\phi(\tau)d\tau+\delta(t)H_m(0).
\end{equation}

Similarly, by substituting \eqref{1.10} into \eqref{1..7}, we find the relation between $R_m(t)$ and $T_m(t)$ as
\begin{equation}\label{B5}
 \sqrt{\pi} 2^m m!R_m(t)= \frac{1}{2}\sum_{k=0}^{m}\frac{m!}{k!(m-k)!}\int_{0}^{t}\sqrt{\pi}2^k k!\left[\left(\frac{2 v_0 \tau}{a(t)}\right)^{m-k}+\left(\frac{-2 v_0 \tau}{a(t)}\right)^{m-k}\right]T_k(t-\tau)\psi(\tau)d\tau .
\end{equation}
\end{widetext}

Equations \eqref{1.11} and \eqref{1.12} can be got in Laplace space by taking Laplace transform on \eqref{B4} and \eqref{B5} with the help of convolution theorem.

\end{appendix}

\bibliography{ref}

\end{document}